\title{Hecke algebraic approach to the reflection equation for spin chains}
\author{A Doikou \\ {{\footnotesize Mathematics, University of York, 
Heslington, York YO10 5DD, UK.}} \\
\\ P P Martin  \\ \myaddress}\date{}
\documentclass[12pt]{article}
\usepackage{epic}
\usepackage{eepic}
\usepackage{graphicx}
\usepackage{latexsym}
\usepackage{amssymb}
\usepackage[all]{xy}
\providecommand{\noglossaryignore}[1]{}
\newcommand{\globalglossaryentry}[3]{\makebox[1.5in][l]{\tt $\backslash${#1}} 
\makebox[1.1in][l]{{$#2$}} \makebox[2.5in][l]{{#3}}\newline} 
\newcommand{\newcommandabbreviation}[3]{\newcommand{#1}{#2}%
\noglossaryignore{\globalglossaryentry{#3}{#2}{}}}
\newcommand{\renewcommandabbreviation}[3]{\renewcommand{#1}{#2}%
\noglossaryignore{\globalglossaryentry{#3}{#2}{}}}
\newcommand{\newcommandmacro}[4]{\newcommand{#1}{#2}%
\noglossaryignore{\globalglossaryentry{#3}{#2}{#4}}}
\newcommand{\gge}[3]{\noglossaryignore{\globalglossaryentry{#1}{#2}{#3}}}
\newcommand{\myaddress}%
{\parbox{3in}{\footnotesize \begin{center} 
Mathematics Department, City University, \\  
Northampton Square, London EC1V 0HB, UK.\end{center}}}
    
\newcounter{minidef}[section]

\newcounter{minicapt}

\newtheorem{de}{Definition}     \newtheorem{pr}{Proposition}

\noglossaryignore{GREEK ETC.\newline}
\newcommandabbreviation{\e}{\epsilon}{e}        
\newcommandabbreviation{\lam}{\lambda}{lam}  
\newcommandabbreviation{\la}{\langle}{la}        
\newcommandabbreviation{\ran}{\rangle}{ran}
\newcommandabbreviation{\ha}{\#}{ha}             
\newcommandabbreviation{\rmap}{\rightarrow}{rmap}
\newcommandabbreviation{\aaa}{\alpha}{aaa}        
\newcommandabbreviation{\ab}{\alpha,\beta}{ab}
\newcommandabbreviation{\aab}{a(\ab )}{aab}       
\noglossaryignore{\newline RINGS\newline}
\newcommandabbreviation{\HH}{H \!\!\! I}{HH}               
\newcommandabbreviation{\C}{\mathbb C}{C}
\newcommandabbreviation{\N}{\mathbb N}{N}   
\newcommandabbreviation{\Z}{\mathbb Z}{Z}      
\renewcommandabbreviation{\Re}{\mathbb R}{Re}
\newcommandabbreviation{\R}{{\mathbb R}}{R}
\newcommandabbreviation{\Q}{\mathbb Q }{Q}
\renewcommandabbreviation{\H}{\mathbb H }{H}
\noglossaryignore{\newline SYMMETRIC GROUP\newline}
\def\Sym(#1){\Sigma(#1)}                   
\gge{Sym(-)}{\Sym(-)}{symmetric group on - objects}
\def\Sy(#1){\Sigma_{#1}}                   
\gge{Sy(-)}{\Sy(-)}{symmetric group irreducible -}
\def\sym(#1){\mbox{\LARGE s}(#1)}        
\gge{sym(-)}{\sym(-)}{symmetric group on - objects (variant)}
\def\sy(#1){\mbox{\LARGE s}({#1})}        
\gge{sy(-)}{\sy(-)}{symmetric group irreducible - (variant)}
\newcommandmacro{\cs}{\C \, \sy(n)}{cs}{symmetric group algebra over $\C$}
\noglossaryignore{\newline PARTITIONS/SETS\newline}
\newcommand{\Nset}[1]{\underline{#1}}
\gge{Nset\{-\}}{\Nset{-}}{set of natural numbers to -} 
\def\nset(#1){ \{ #1 \}_{ \underline{n} }} 
\gge{nset(-)}{\nset(-)}{a set $-\times\Nset$}
\def\ul(#1){_{\underline{#1}}}             
\gge{ul(-)}{{}\ul(-)}{subscript underline -}
\def\Ee(#1){{\bf E}_{#1}}                  
\gge{Ee(-)}{\Ee(-)}{set of equivalence relations on set -}
\def\Eee(#1){{\bf E}_{\{ #1 \}_{\underline{n}}}}   
\gge{Eee(-)}{\Eee(-)}{ditto for nset}
\def\Een(#1,#2){{\bf E}_{\{ #1 \}_{\underline{#2}}}}   
\def\Ssn(#1,#2){{\bf S}_{\{ #1 \}_{\underline{#2}}}}   
\def\Ss(#1){{\bf S}_{#1}}                  
\def\Sss(#1){{\bf S}_{\{ #1 \}_{\underline{n}}}}   
\def\bbc(#1){((\beta_1)(\beta_2)...(\beta_{#1}))}      
\newcommandmacro{\Ln}{{\Gamma}^{n}}{Ln}{large index set}
\newcommandmacro{\LnQ}{{\Gamma}^{n}_Q}{LnQ}{index set}
\newcommandmacro{\Zz}{\zeta}{Zz}{`shape' function}
\noglossaryignore{\newline PARTITION ALGEBRA\newline}
\def\ka(#1){\kappa_{#1}}                   
\def\Sm(#1){\Sigma_{#1}}                   
\newcommandmacro{\com}{\bullet}{com}{bullet composition}
\newcommandmacro{\enm}{\; e^n(\! m\! ) \;}{enm}{product of idempotents}
\def\Ai(#1){ A^{ #1 \cdot } }              
\def\Aij(#1,#2){ A^{ #1  #2 } }            
\newcommandmacro{\One}{\mbox{\bf $1 \!\!\! 1$}}{One}{algebra unit 1}
\newcommandmacro{\Bp}{B_p}{Bp}{partition basis}
\def\Bb(#1){B_p[#1]}                       
\def\Pp(#1){P_n[#1]}                       
\def\Ps(#1){P_n[#1] \! /}                  
\newcommandmacro{\Ph}{\hat{P}}{Ph}{P hat  algebra}
\def\Is(#1){\sim^{#1}}                     
\noglossaryignore{\newline MODULES\newline}
\def\Wm(#1){{\cal S}_{#1}}                 
\gge{Wm(-)}{\Wm(-)}{Weyl module with index -}
\def\wm(#1,#2){{}_{#1}{\cal S}_{#2}}       
\gge{wm(-1,-)}{\wm(-1,-)}{Weyl module with index -}
\def\Ind(#1,#2,#3){\mbox{Ind}_{#1}^{#2}#3} 
\gge{Ind(-1,-2,-)}{\Ind(-1,-2,-)}{induction}
\def\Res(#1,#2,#3){\mbox{Res}_{#1}^{#2}#3} 
\gge{Res(-1,-2,-)}{\Res(-1,-2,-)}{restriction}
\newcommandabbreviation{\weyl}{standard}{weyl}
\newcommandabbreviation{\mod}{\mbox{mod}}{mod}
\newcommandabbreviation{\head}{\mbox{head }}{head}
\newcommandabbreviation{\Weyl}{Weyl}{Weyl}
\def\SS(#1){{\cal S}_{#1}}                 
\gge{SS(-)}{\SS(-)}{Specht/Weyl module index -}
\def\LL(#1){{\cal L}_{#1}}                 
\gge{LL(-)}{\LL(-)}{simple module index -}
\noglossaryignore{\newline FUNCTORS/MAPS\newline}
\newcommandmacro{\Gg}{{\cal G}}{Gg}{G Functor}
\newcommandmacro{\Fg}{{\cal F}}{Fg}{F Functor}
\newcommandmacro{\ra}{\rightarrow}{ra}{}
\def\ses(#1,#2,#3){0\ra #1 \ra #2 \ra #3 \ra 0}   
\gge{ses(1,2,-)}{\ses(1,2,-)}{\hspace{.5in} short exact sequence}
\def\starr(#1){ \stackrel{ #1 }{\longrightarrow} }
\gge{starr(-)}{\starr(-)}{}
\newcommandmacro{\doublerightarrow}{\; -\!\!\! -\!\!\!\!\!\! \gg \;}
{doublerightarrow}{}
\noglossaryignore{\newline PARTITION ALGEBRA MAPS\newline}
\newcommandmacro{\smap}{s}{smap}{`inclusion' map}
\newcommandmacro{\tmap}{t}{tmap}{$ P_n -> S_n$}
\newcommandmacro{\pmap}{\psi}{pmap}{$ S_n -> P_n $}
\noglossaryignore{\newline MISC.\newline}
\def\Amap(#1){{\cal A}_{#1}}               
\gge{Amap(-)}{\Amap(-)}{}
\def\Rr(#1){R_{#1}}                        
\gge{Rr(-)}{\Rr(-)}{restriction of E}
\def\Cr(#1){C_{#1}}                        
\gge{Cr(-)}{\Cr(-)}{restriction of E to N}
\newcommandmacro{\Tm}{{\cal T}}{Tm}{Transfer Matrix}
\def\On(#1){{\cal I}_{#1}}
\gge{On(-)}{\On(-)}{}
\newcommandmacro{\UU}{\underline{\sqcup}}{UU}{}  
\newcommandmacro{\UUU}{\sqcup}{UUU}{}  
\newcommandmacro{\Vq}{V_Q^{\otimes n}}{Vq}{Potts config. space}
\def\bs(#1,#2){\mbox{{\Large $\ast$}}^{#1}_{#2}}  
\gge{bs(-,-)}{\bs(-,-)}{general plumbing multiplier}
\newcommand{\ignore}[1]{}
\gge{ignore\{-\}}{\ignore{-}}{ignore argument!}
\def\choo(#1,#2){ \left( \begin{array}{c} #1 \\ #2 \end{array} \right) } 
\gge{choo(-1,-)}{\choo(-1,-)}{choose}
\newcommand{\Qed}{$\Box$}
\gge{Qed}{\mbox{\Qed}}{QED}
\def\staq(#1){\stackrel{#1}{=}}            
\gge{staq(-)}{\staq(-)}{}
\def\stam(#1){\stackrel{#1}{\rightarrow}}  
\gge{stam(-)}{\stam(-)}{}
\def\mat{ \left( \begin{array} }    
\def\tam{ \end{array}  \right) }
\gge{mat/tam}{...}{matrix delimiters}
\newcommand{\beq}{\begin{equation} }
\def\eql(#1){ \begin{equation} \label{#1} 
}
\newcommand{\eq}{\end{equation} }
\def\eqal(#1){\begin{eqnarray} \label{#1} }
\def\eqa{\end{eqnarray} }
\def\lab(#1){\label{#1}
}
\def\prl(#1){ \begin{pr} \label{#1} 
}
\def\del(#1){ \begin{de} \label{#1} 
}
\gge{smeq\{-\}}{...}{small equation}
\gge{fneq\{-\}}{...}{very small equation}
\noglossaryignore{\newline HECKE/BLOB\newline}
\newcommandmacro{\Hnq}{H_n(q)}{Hnq}{ * freestanding symbol}
\newcommandmacro{\Hn}{H_n}{Hn}{      *-mod etc.}
\newcommandmacro{\A}{{\cal A}}{A}{}
\newcommandmacro{\Cwts}{C}{Cwts}{}
\newcommandmacro{\CA}{{\cal A}}{CA}{}

\newcommandmacro{\calA}{{\cal A}}{calA}{}
\newcommandmacro{\modi}{\mbox{Mod} }{modi}{was mod not modi!}
\newcommandmacro{\Wgen}{{\Bbb S}}{Wgen}{}
\def\ol(#1){\overline{#1}}
\newcommandmacro{\st}{\mbox{St}}{st}{}
\def\CMult(#1,#2){(#1:#2)}
\def\CM(#1,#2){( #1 : #2 )}
\def\FMult#1,#2{(#1:#2)}
\def\CF#1,#2{(#1:#2)}

\newcommandmacro{\Top}{\mbox{Top}}{Top}{}
\newcommandmacro{\Soc}{\mbox{Soc}}{Soc}{}
\newcommandmacro{\Head}{\mbox{Head}}{Head}{}
\newcommandmacro{\Filt}{{\cal F}}{Filt}{}
\newcommandmacro{\Mod}{\mbox{mod}}{Mod}{}
\newcommandmacro{\Resi}{\mbox{Res }}{Resi}{was without i!}
\newcommandmacro{\Indi}{\mbox{Ind }}{Indi}{was without i!}
\def\RR(#1,#2){R^{#1}_{#2}}   
\def\TT(#1,#2){T^{#1}_{#2}}   

\newcommandmacro{\Ann}{\mbox{Ann}}{Ann}{}
\newcommandmacro{\Cen}{\mbox{Cen}}{Cen}{}
\newcommandmacro{\End}{\mbox{End}}{End}{}
\newcommandabbreviation{\semisimple}{semisimple}{semisimple}
\newcommandabbreviation{\Bratteli}{Bratteli}{Bratteli}
\newcommandabbreviation{\JBC}{Jones Basic Construction}{JBC}
\newcommandabbreviation{\pa}{partition algebra}{pa}
\newcommandabbreviation{\TM}{transfer matrix}{TM}
\newcommandabbreviation{\PM}{Potts model}{PM}
\newcommandabbreviation{\QSC}{quantum spin chain}{QSC}
\newcommandabbreviation{\Hamiltonian}{Hamiltonian}{Hamiltonian}
\newcommandabbreviation{\YS}{Young symmetrizer}{YS}

\newcommand{\be}{\begin{eqnarray}}
\newcommand{\eeq}{\end{eqnarray}}
\newcommand{\non}{\nonumber}

\newcommand{\XR}{\ensuremath{\mathcal{R}}}
\newcommand{\tr}{\mathop{\rm tr}\nolimits}

\newcommand{\TL}{Temperley--Lieb}
\newcommand{\YB}{Yang--Baxter}
\newcommand{\rank}{N}
\newcommand{\sites}{n}
\newcommand{\lateral}{lateral}

\newcommand{\braid}[1]{{\cal B}_{#1}}    
\newcommand{\braido}[1]{{\cal B}^{\circ}_{#1}} 
\newcommand{\Artin}[1]{{\cal A}_{#1}}

\newcommand{\xfiginps}[3]{\begin{figure}\includegraphics[width=#3in]{xfig/#1.eps}%
\caption{\label{#1} #2}\end{figure}}
\newcommand{\affine}[1]{\hat{#1}}
\newcommand{\twist}{\tau}
\newcommand{\squash}{\sigma} \newcommand{\sm}{l}
\newcommand{\glue}{\iota}
\newcommand{\cable}{\gamma}
\newcommand{\calR}{{\cal R}}

\newcommand{\fk}{FringKoberle94a,FringKoberle94b}

\newcommand{\DVGRi}{deVegaGonzalezruiz94c}

\begin{document} \maketitle
\newcommand{\ygnore}[1]{}
 \newcommand{\ignoreifnotdraft}[1]{\ygnore{#1}}
\ignoreifnotdraft{
\pagestyle{myheadings} \markboth{Draft}{\today}
}

\ignoreifnotdraft{
\noindent
\begin{tabular}{l}
  {\tiny \filename (Draft)} \hspace{4.4in} Jan 1997 \\ \hline
\end{tabular}
}
\begin{abstract}
We use the structural similarity of certain Coxeter Artin Systems to
the Yang--Baxter and Reflection Equations to convert representations
of these systems into new solutions of the Reflection Equation. 
We construct certain Bethe ansatz states for these solutions, using a
parameterisation suggested by abstract representation theory. 
\end{abstract}
\ignoreifnotdraft{ \newpage } 
\newcommand{\AR}{A}
\newcommand{\paula}{\sigma}
\newcommand{\UUUU}{{\mathcal U}}
\newcommand{\Uqsl}{U_{q}sl_}
\section{Introduction and review}
There has been much interest recently in the role of boundaries in 
integrable systems, both from the point of view of
critical phenomena (see for example \cite{AffleckOshikawaSaleur98} and
references therein), and integrability \cite{GhoshalZamolodchikov94a}.
There has also been considerable progress in constructing
representations of affine
Hecke algebras \cite{DipperJamesMathas99,KashiwaraMiwaStern95}
with global (i.e. quasi--thermodynamic) limits 
\cite{MartinWoodcockLevy00,MartinWoodcock01pre}.
In this paper we apply this algebraic technology to the boundary
$R$--matrix problem, in a way analogous to the use by many authors of
the ordinary Hecke algebra in solving the Yang--Baxter equations
(see \cite{Martin91,Henkel99} for reviews).

We start by briefly reviewing the standard $R$--matrix formulation of
the \YB\ equation (YBE) in the context of spin chains, 
and the Hecke/\TL\ algebraic variant of this formulation. 
We then generalise to $K$--matrices
and boundary YBE --- i.e. to the reflection equation (RE)
\cite{Cherednik84,Sklyanin88}. 
In \S\ref{Sbraids} we discuss the
algebraic structures with roles {\em analogous} to the ordinary Hecke and \TL\
algebras in the boundary case, 
and give a number of constructions for representations of such
algebras, which representations provide candidates for solutions to RE.
In \S\ref{SSolutions} we show that the resultant `blob algebra' $b_n$ indeed
provides new (and well parameterized) solutions to RE. 
Finally  we look at
the Bethe ansatz for some intriguing `spin--chain--like'
representations of this algebra.

The parallels with the ordinary closed boundary $U_qsl_2$--invariant
spin chain case are strong, 
but the symmetry algebra is {\em not} always $U_qsl_2$. This
raises some very interesting questions for further study.
The representation theory of $b_n$ has parallels with that of the
Virasoro algebras arising in conformal field theory, and Bethe ansatz
may provide a mechanism for investigating this
(cf. \cite{KooSaleur93,RochaCaridi84,Henkel99,MartinWoodcock2000}).


Fix integers $\rank >0$ and $\sites >> 0$, and  
let $V$ be complex $\rank$--space.
Write $V^{\sites}=\otimes_{i=1}^{\sites} V$. For $N=2$ the Pauli
$\paula$--matrices, and indeed $\Uqsl2$, act naturally on $V$, and
$V^{\sites}$ is the underlying space of the $\sites$--site XXZ
model. Define
$H^{\rank}_{\sites}(q)=\End_{\Uqsl{\rank}}(V^{\sites})$. 
The ordinary \TL\ algebra $T_{\sites}(q)$ (see later) is isomorphic to 
$H^2_{\sites}(q)$.  

\subsection{$R$--matrices}
Define ${\cal P}$ to act on $V\otimes V$ by ${\cal P} x \otimes y = y \otimes x$.
If $\AR$ is any matrix acting on 
$V^m=\otimes_{i=1}^m V$, and $ i_1,...,i_m \leq n$ distinct natural numbers,
then (in `$R$--index notation')
$\AR_{i_1...i_m}$ acts on $V^{\sites}$ by embedding
the $\AR$ action onto the $i_1^{th}$... $i_m^{th}$ factors $V$.
Eg.,
${\cal P}_{12}={\cal P}_{21}$ and
${\cal P}_{12}{\cal P}_{13}{\cal P}_{12}={\cal P}_{23}$.
Dually, if $T$ is a matrix acting on $V\otimes V^n$ (with factors
indexed from $0,1,...,n$) then $T_0$
is $T$ regarded as an
$\rank^{n} \times \rank^{n}$--matrix--valued
$\rank \times \rank$--matrix in the obvious way.
Generalising this (for a moment) 
so that $T_i$ is $T$ expanded with respect to the
$i^{th}$ factor then
$\tr_i(T)=\tr(T_i)$, the trace (we may also write this as $\tr_i(T_i)$);
and $T^{t_i}=(T_i)^t$, the transpose.

An (adjoint) {\em $R(\lambda)$--matrix} is a matrix acting on $V^2$ which
solves the Yang--Baxter equation in the ($R$--index) form
\cite{KorepinIzerginBogoliubov93}
\be
R_{12}(\lambda-\lambda')\ R_{13}(\lambda)\ R_{23}(\lambda') =
R_{23}(\lambda')\ R_{13}(\lambda)\ R_{12}(\lambda-\lambda') .
\label{YB0} \eeq
We also require {\em unitarity}:
\be R_{12}(\lambda) R_{21}(-\lambda) \propto 1 \,
\label{property10} \eeq
(NB, $R_{21}(\lambda) = {\cal P}_{12} R_{12}(\lambda) {\cal P}_{12} $);
$R_{21}(\lambda) = R_{12}(\lambda)^{t_{1} t_{2}}$;
and
\cite{ReshetikhinSemenovTianShansky90}
that there exist $ M = M^{t}$ and $\rho$ such that 
\be R_{12}(\lambda)^{t_{1}} M_{1}
R_{12}(-\lambda - 2\rho)^{t_{2}} M_{1}^{-1} \propto 1 \,,
\label{property20} \eeq
\be \left[ M_{1} M_{2}
\,, R_{12}(\lambda) \right] = 0 \,.
 \label{property30}
 \eeq

Given such an $R(\lambda)$--matrix,
introduce {\em monodromy matrix}
\cite{FaddeevTakhtajan79,FaddeevTakhtajan81}
\be T(\lambda) = R_{0 \sites}(\lambda)
\cdots R_{0 1}(\lambda) \,. \eeq
NB, this acts on
$V\otimes V^n = V_0 \otimes V_1 \otimes V_2 \ldots V_n$.
Spaces $V_{i}$ ($i>0$) are called `quantum';
space $V_0$ is called `\lateral' or `auxiliary'.
One often makes manifest just the
\lateral\ space subscript: $T(\lambda)=T_{0}(\lambda)$.
The YBE implies
\be
R_{00'}(\lambda-\lambda')\ T_{0}(\lambda)\ T_{0'}(\lambda') =
T_{0'}(\lambda')\ T_{0}(\lambda)\ R_{00'}(\lambda-\lambda') \,.
\label{fundamental0} \eeq
It will be convenient in what follows to have in mind a pictorial
realisation of the verification of equation(\ref{fundamental0}). 
One represents the YBE itself as in figure~\ref{braidYB}.
\xfiginps{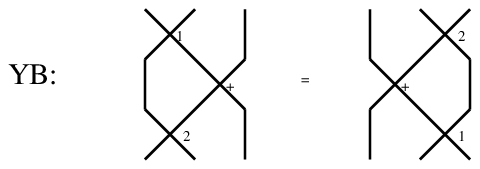}{Pictorial realisation of the YBE generalising the
permutation diagram realisation of the symmetric group. Here a
crossing labelled by 1 (resp. 2, $+$) represents $R_{ij}(\theta_1)$ 
(resp. $R_{ij}(\theta_2)$, 
$R_{ij}(\theta_1 +\theta_2)$), 
and $\theta_1=\lambda-\lambda'$, $\theta_2=\lambda$.}{4}
\xfiginps{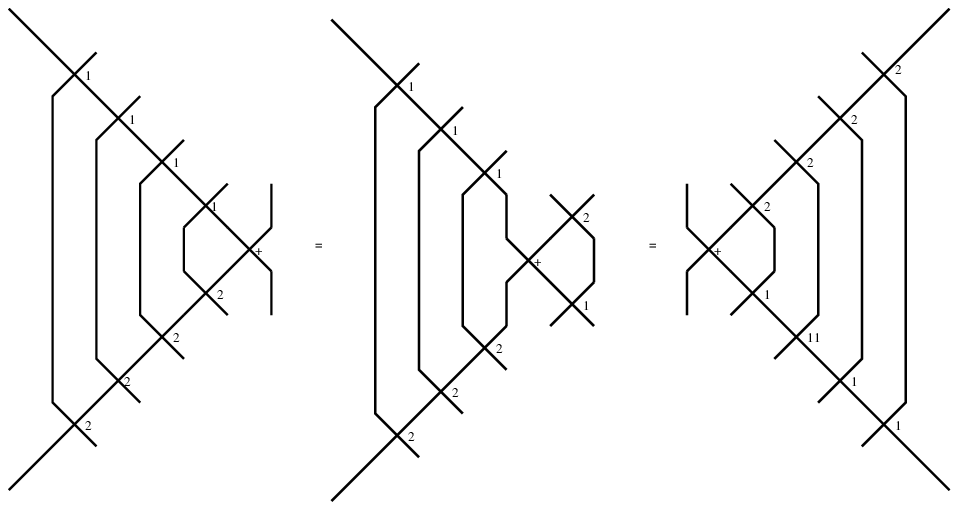}{Application of the YBE to verify commutation.}{6}
The identity then follows by repeated application of 
the YBE as in figure~\ref{braidYB02}.


The {\em closed chain transfer matrix} is 
\be
t(\lambda) = \tr_{0} T_{0}(\lambda) \, \label{transfer} \eeq
By virtue of (\ref{fundamental0}) and the existence of inverse of
$R(\lambda)$ this obeys
\be \left[ t(\lambda)\,, t(\lambda') \right] = 0 . 
\eeq

Example: with $N=2$ the 
XXZ model with anisotropy parameter $\mu \ge 0$ 
has \cite{deVega89}
\be \label{XXZR}
R(\lambda)=\left( \begin{array}{cccc}
    a(\lambda)                                  \\
    &         b(\lambda) & c_{+}  (\lambda)                  \\
    &         c_{-}(\lambda)          & b(\lambda)           \\
    &                    &           & a(\lambda)
\end{array} \right) \label{xxz} \,
\eeq
 where
 \be a(\lambda) &=& \sinh \left( \mu (\lambda + i)
\right)
\non  \\
b(\lambda) &=& \sinh (\mu \lambda) \non  \\
c_{\pm}(\lambda) &=& \sinh (i \mu)e^{\pm \mu \lambda} \,
 \eeq
(also known as $A_{1}^{(1)}$ case, 
by an association with the $A_{1}^{(1)}$ affine Lie algebra). 
This $R$--matrix obeys (\ref{property20}) and (\ref{property30}) with
\cite{deVegaGonzalezruiz94a,deVegaGonzalezruiz93}
 \be M_{j k} = \delta_{j k} e^{i \mu (3 - 2 j) }\,,
\qquad \ \rho = i \,.
 \eeq





\subsection{$R$--matrices and the TL algebraic method}
Given an $R$--matrix, set
\be
\check R_{i i+1}(\lambda)
= {\cal P}_{i i+1} R_{i i+1}(\lambda) = R_{i+1 i}(\lambda){\cal P}_{i i+1} .
 \eeq
Premultiplying (\ref{YB0}) by ${\cal P}_{23}{\cal P}_{12}{\cal
  P}_{23}$ we get
\be \check R_{12}(\lambda-\lambda')\ \check
R_{23}(\lambda)\ \check R_{12}(\lambda')
= \check
R_{23}(\lambda')\ \check R_{12}(\lambda)\ \check
R_{23}(\lambda-\lambda')  \label{YB03} \eeq
What is deep about (\ref{YB0}) is the construction of commuting
transfer matrices, and this is not restricted to, and may be
abstracted away from, the $V^{\sites}$ setting. One introduces
abstract operators $ \check R_{i}(\lambda) $ (not in $R$--index notation)
obeying
\be \check R_{i}(\lambda-\lambda')\
\check R_{i+1}(\lambda)\
\check R_{i}(\lambda')
= \check R_{i+1}(\lambda')\
\check R_{i}(\lambda)\
\check R_{i+1}(\lambda-\lambda')  \label{YB02} \eeq
and
\be
\check R_{i}(\lambda)\
\check R_{j}(\lambda')
= \check R_{j}(\lambda')\
\check R_{i}(\lambda)\
\qquad i-j>1 \,. \label{YB02commutation} \eeq
This  is called the {\em Hecke algebraic form} of the YBE.
It will be evident that every $R$--matrix gives a solution to these
equations via the substitution
$ \check R_{i}(\lambda) \mapsto  \check R_{i i+1}(\lambda) $.

The abstract Temperley--Lieb algebra $T_{\sites}(q)$ is generated by the unit
element and elements $U_{1}, \ldots, U_{\sites -1}$ satisfying the following
relations
\cite{TemperleyLieb71,Baxter82}
 \be
 U_{i}U_{i} &=&-(q+q^{-1}) U_{i},
          \qquad q = e^{i \mu} \non\\
U_{i}U_{i \pm 1}U_{i} &=& U_{i} \non\\
\left[U_{i}, U_{j} \right] &=& 0\, \qquad i-j
>1.
\label{algebra}
 \eeq
Let $\rank=2$, and $V^{\sites}$ the corresponding
tensor space with action of $U_qsl_2$ \cite{KulishSklyanin91,Martin91}.
Set
\be \calR (U_{i}) = \calR_q (U_{i}) 
&=& \paula^+_i \paula^-_{i+1}+\paula^-_i \paula^+_{i+1}
    +\frac{q+q^{-1}}{1} \left( \paula^z_i \paula^z_{i+1} - \frac{1}{4}
    \right)
    +\frac{q-q^{-1}}{2} \left( -\paula^z_i +\paula^z_{i+1} \right)
\non \\ \label{rrrep}
&=& 1\otimes \ldots \otimes \UUUU  \otimes \ldots \otimes 1
 \eeq
where \be \UUUU =
 \left(
\begin{array}{cccc}
    0                                  \\
    &         -e^{i\mu} &  1                   \\
    &         1  & -e^{-i\mu}           \\
    &            &           & 0
\end{array} \right) \label{u} \eeq
(i.e. the nontrivial part is a $4 \times 4$ matrix acting on $V_{i} \otimes
V_{i+1}$, so ${\cal R}(U_{i})=\UUUU_{i i+1}$ in $R$--index notation).

\prl(faith)\mbox{{\rm \cite{Martin92}}}
The matrices $ {\cal R}(U_{i})$ define a representation of
$T_{\sites}(q)$ which is
(i) faithful; and
(ii) commutes with  the action of $U_qsl_2$ on $V^{\sites}$.
\end{pr}
For the XXZ $R$--matrix of equation~(\ref{XXZR}) we find
\be
 \check R_{i i+1}(\lambda)= \sinh(\mu(\lambda +i))1 + \sinh(\mu
 \lambda) {\cal R}( U_{i} ).
\label{R}
 \eeq
Thus ${\cal R}$ gives a solution to (\ref{YB03}) and hence to (\ref{YB0}).
Since ${\cal R}$ is faithful,
any representation of $T_{\sites}(q)$ would give a solution to (\ref{YB02}).
We say $T_{\sites}(q)$ gives a {\em meta--solution}.

\subsection{$K$--matrices}

Given an $R$--matrix, 
a {\em $K(\lambda)$--matrix} acts on $V$
and obeys the reflection equation \cite{Cherednik84}:
\be R_{12}(\lambda_{1}-\lambda_{2})\
K_{1}(\lambda_{1})\
R_{21}(\lambda_{1}+\lambda_{2})\ K_{2}(\lambda_{2}) \non \\
= K_{2}(\lambda_{2})\ R_{12}(\lambda_{1}+\lambda_{2})\
K_{1}(\lambda_{1})\ R_{21}(\lambda_{1}-\lambda_{2}) \,.
\label{boundaryYB0B} \eeq
We require $K(0) = 1$ and $K(\lambda) K(-\lambda) \propto 1$.
Using this one may construct commuting {\em open} boundary transfer
matrices and solve corresponding Bethe ansatz equations
\cite{Sklyanin88}.


A suitable transfer matrix $t(\lambda)$ for an open chain
of $\sites$ spins is 
\cite{Sklyanin88,KulishSklyanin91,MezincescuNepomechie91a}
\be t(\lambda) = \tr_{0} M_{0}\ K_{0}^{+}(-\lambda-\rho)^{t}\
T_{0}(\lambda)\  K^{-}_{0}(\lambda)\ \hat T_{0}(\lambda)\,,
\label{transfer10} \eeq 
where 
\be \hat
T_{0}(\lambda) = R_{10}(\lambda) \cdots  R_{\sites 0}(\lambda) \,,
\label{hatmonodromy0} \eeq 
$K^{-}(\lambda) = K(\lambda)$
where the $K(\lambda)$ is a solution of the reflection equation, 
and $K^+$ satisfies an equation similar to (\ref{boundaryYB0B}) 
\cite{DoikouNepomechie98}
(we can and will set $K^+=1$ without significant loss of generality).


Following Sklyanin \cite{Sklyanin88} define
\be {\cal T}(\lambda) = T_{0}(\lambda)\
K^{-}_{0}(\lambda)\ \hat T_{0}(\lambda)\,, \label{monodromyo} \eeq
which satisfies 
 \be R_{12}(\lambda_{1}-\lambda_{2})\ {\cal
T}_{1}(\lambda_{1})\
R_{21}(\lambda_{1}+\lambda_{2})\ {\cal T}_{2}(\lambda_{2}) \non \\
= {\cal T}_{2}(\lambda_{2})\ R_{12}(\lambda_{1}+\lambda_{2})\
{\cal T}_{1}(\lambda_{1})\ R_{21}(\lambda_{1}-\lambda_{2}) \,.
\label{bound} \eeq
We may again use the pictorial representation to see this. 
Following figure~\ref{braidYB} the picture for the 
reflection equation (\ref{boundaryYB0B}) is as in figure~\ref{braidRE1}. 
\xfiginps{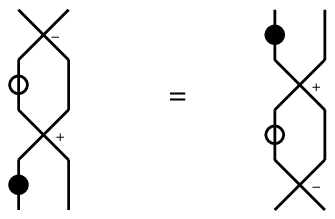}{Pictorial realisation of the RE}{3}
In this realisation the Sklyanin operator appears as in
figure~\ref{braidRE2}. 
\xfiginps{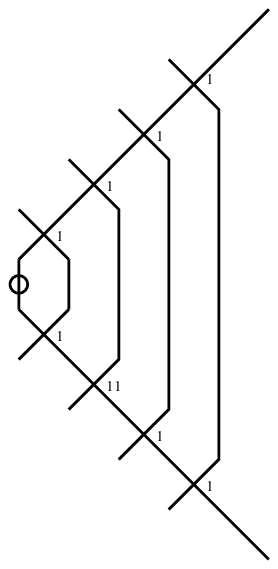}{The Sklyanin operator ${\cal T}(\lambda)$.}{2}
The identity (\ref{bound}) follows in the manner of figure~\ref{braidRE3}.
\xfiginps{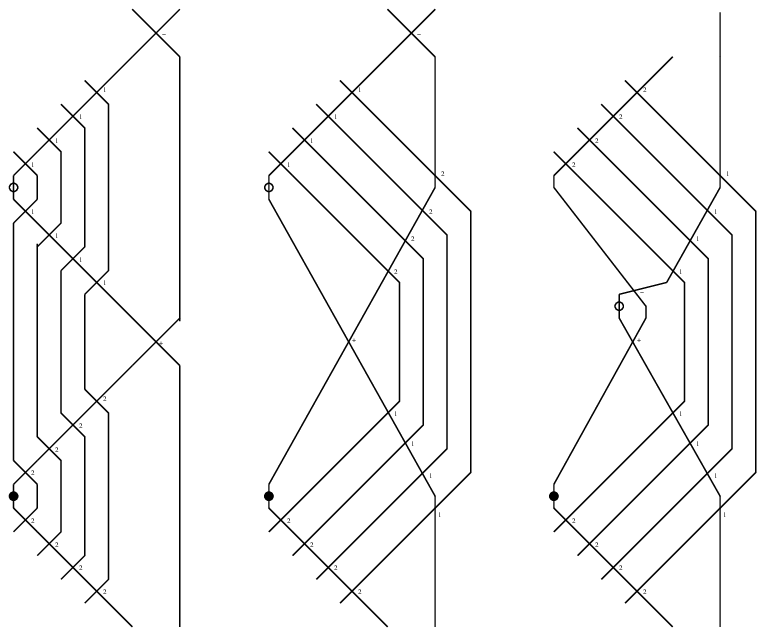}{First steps in verification of commutation. Step 1
is an application of YBE as in figure~\ref{braidYB02}. Step 2 is
similar. At this point the left hand side of RE has appeared in
the picture. One applies RE to it and then completes the
manipulation by further applications of YBE.}{6}

The transfer matrix also obeys 
\be \left[ t(\lambda)\,, t(\lambda')
\right] = 0 \,. \label{commutativity4} \eeq

Consider the XXZ/$A_{1}^{(1)}$ $R$--matrix as before. 
For $K=1$, 
\be \left[ t(\lambda), g \right] = 0 \,
\label{second0} \eeq 
where $g$ is the usual $U_qsl_2$ action 
\cite{PasquierSaleur90,MezincescuNepomechie92,MezincescuNepomechie91c,Sklyanin88}.
The symmetry for the
general diagonal $K$ is more complicated (see e.g. \cite{DoikouNepomechie98}).


In the \TL\ notation the RE is
\be \check R_{1}(\lambda_{1} \! - \! \lambda_{2})\
\check K_{}(\lambda_{1})\
\check R_{1}(\lambda_{1}\! +\! \lambda_{2})\
\check K_{}(\lambda_{2})
= \check K_{}(\lambda_{2})\ \check R_{1}(\lambda_{1} \! + \!
\lambda_{2})\ \check K_{}(\lambda_{1})\ \check R_{1}(\lambda_{1}
\! - \! \lambda_{2}) \label{boundaryYB0Babstract} . \eeq 
As we will now see, this makes it natural to seek solutions among the
affine generalisations of $T_{\sites}(q)$.

\section{
(Affine) braids and Hecke algebras}\label{Sbraids}
Recall that a {\em Coxeter graph} $G$ is any finite undirected graph without
loops (almost everybody's attention is habitually restricted to the
subset of graphs of {\em positive type} \cite[\S2.3]{Humphreys90}). For
given $G$ let
$m(s,s')$ denote the number of edges between vertices $s$ and $s'$.
The {\em Coxeter system} of $G$ is a pair
$(W,S)$ consisting of a group $W$ and a set
$S$ of generators of $W$ labelled by the vertices of $G$,
with relations of the form
\eql(coxeter)
g_s g_{s'} g_s g_{s'} ... = g_{s'} g_s g_{s'} g_s ...
\eq
where the number of factors on each side is $m(s,s')+2$; and
\eql(local1)
g_s^{-1}  = g_s .
\eq
If we relax the set of relations in (\ref{local1}) (and add as
generators the inverse of each $g_s\in S$) we get a
{\em Coxeter Artin  system},
and $W=\Artin{G}$ is an {\em Artin group} \cite{BrieskornSaito72}.
For example,
let $\braid{n}$ denote the ordinary Artin braid
group, the group of composition of finite braidings of $n$ strings running from
the northern to the southern edge of a rectangular frame.
Then $\Artin{A_{n-1}} \cong \braid{n}$.

In case
$G=B_n$
the (non--commuting) relations may be written
\eql(4rel) g_0 g_1 g_0 g_1 = g_1 g_0 g_1 g_0 \eq
\eql(3rel)
g_i g_{i+1} g_i = g_{i+1} g_i g_{i+1} \hspace{2cm} n-1 > i\geq 1 .
\eq
{\em And here is the {\em point} of this excursion}: We will use the structural
similarity of these relations to the
reflection equation (RE) \cite{Cherednik84,Sklyanin88} and
Yang--Baxter equation (YBE) \cite{Baxter82} to develop various
realisations of
$\Artin{B_n}$   
into candidates for solutions to these equations.
There are two parts to this task. Finding quotients of the braid group
in which (\ref{4rel}) and (\ref{3rel}) may be deformed to solve RE and
YBE respectively (see \S\ref{ss2}); and then finding realizations of
these quotients suitable for Bethe ansatz formulation. Our approach to
the latter problem is to borrow from what works in the ordinary case
\cite{Baxter82}. Thus we have to make {\em contact} with the ordinary
case. We do this next.


\subsection{Boundaries, cylinder braids and $\Artin{B_n}$}\label{ss1.1}
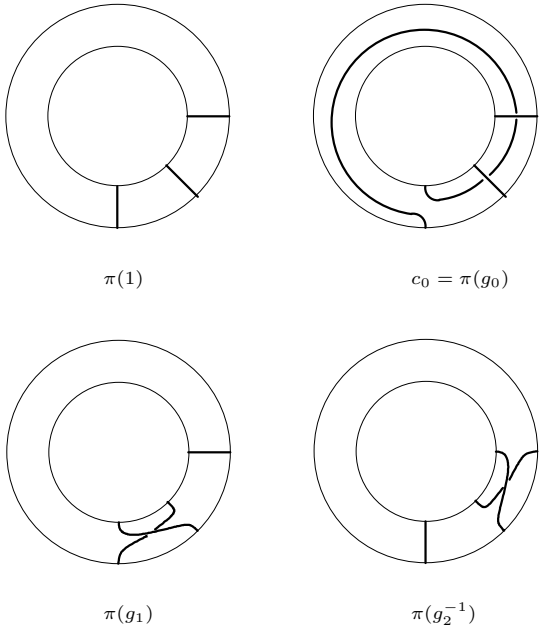
\begin{figure}\setlength{\unitlength}{0.00025000in}
\begingroup\makeatletter\ifx\SetFigFont\undefined%
\gdef\SetFigFont#1#2#3#4#5{%
  \reset@font\fontsize{#1}{#2pt}%
  \fontfamily{#3}\fontseries{#4}\fontshape{#5}%
  \selectfont}%
\fi\endgroup%
{\renewcommand{\dashlinestretch}{30}
\begin{picture}(11430,13330)(0,-10)
\put(2408,10908){\ellipse{4800}{4800}}
\put(2408,10908){\ellipse{3000}{3000}}
\thicklines
\path(2408,9408)(2408,8508)
\path(3458,9858)(4133,9183)
\path(3908,10908)(4808,10908)
\thinlines
\put(9023,3711){\ellipse{3014}{3014}}
\put(9016,3718){\ellipse{4800}{4800}}
\thicklines
\path(9016,2218)(9016,1318)
\path(10523,3711)(10524,3711)(10529,3711)
	(10540,3710)(10557,3708)(10579,3706)
	(10601,3702)(10622,3698)(10641,3693)
	(10658,3687)(10673,3680)(10686,3672)
	(10699,3661)(10707,3654)(10715,3646)
	(10722,3637)(10730,3626)(10738,3615)
	(10745,3602)(10752,3588)(10759,3572)
	(10765,3555)(10771,3536)(10776,3516)
	(10780,3495)(10784,3472)(10787,3448)
	(10789,3423)(10790,3397)(10790,3370)
	(10789,3342)(10787,3313)(10785,3282)
	(10782,3259)(10779,3235)(10775,3210)
	(10771,3185)(10767,3157)(10762,3129)
	(10756,3100)(10750,3070)(10744,3040)
	(10737,3008)(10730,2976)(10723,2943)
	(10715,2910)(10708,2877)(10700,2844)
	(10692,2811)(10684,2779)(10676,2748)
	(10669,2717)(10661,2687)(10654,2657)
	(10647,2629)(10640,2603)(10634,2577)
	(10628,2552)(10622,2529)(10617,2507)
	(10612,2486)(10605,2455)(10599,2426)
	(10593,2399)(10588,2373)(10584,2349)
	(10581,2326)(10578,2304)(10576,2285)
	(10575,2266)(10574,2249)(10574,2234)
	(10575,2220)(10576,2208)(10578,2196)
	(10580,2186)(10583,2177)(10586,2168)
	(10589,2160)(10595,2147)(10602,2135)
	(10611,2122)(10621,2109)(10634,2093)
	(10648,2076)(10663,2059)(10678,2042)
	(10691,2029)(10699,2020)(10702,2017)(10703,2016)
\path(10086,2660)(10087,2659)(10090,2655)
	(10099,2646)(10112,2633)(10128,2616)
	(10145,2599)(10162,2584)(10176,2570)
	(10189,2559)(10201,2550)(10211,2542)
	(10222,2536)(10229,2532)(10237,2528)
	(10245,2526)(10253,2523)(10262,2522)
	(10271,2522)(10281,2522)(10291,2524)
	(10302,2527)(10313,2531)(10324,2537)
	(10335,2543)(10347,2551)(10359,2560)
	(10371,2570)(10383,2581)(10394,2591)
	(10405,2602)(10417,2615)(10430,2629)
	(10444,2646)(10460,2664)(10477,2685)
	(10496,2708)(10516,2733)(10538,2760)
	(10561,2789)(10584,2817)(10606,2845)
	(10626,2871)(10643,2893)(10656,2910)
	(10665,2921)(10671,2928)(10673,2931)
\path(10816,3111)(10817,3114)(10820,3119)
	(10826,3129)(10834,3145)(10844,3166)
	(10858,3191)(10874,3221)(10891,3253)
	(10909,3287)(10928,3322)(10947,3355)
	(10965,3388)(10982,3419)(10999,3447)
	(11014,3474)(11029,3498)(11043,3519)
	(11056,3539)(11068,3556)(11080,3572)
	(11092,3587)(11103,3600)(11115,3612)
	(11132,3629)(11149,3643)(11167,3655)
	(11186,3666)(11207,3676)(11229,3684)
	(11254,3691)(11281,3698)(11308,3704)
	(11335,3708)(11360,3712)(11380,3715)
	(11395,3717)(11404,3718)(11407,3718)(11408,3718)
\thinlines
\put(2436,3692){\ellipse{3014}{3014}}
\put(2429,3699){\ellipse{4800}{4800}}
\thicklines
\path(3929,3699)(4829,3699)
\path(2436,2192)(2436,2191)(2436,2186)
	(2437,2175)(2439,2158)(2441,2136)
	(2445,2114)(2449,2093)(2454,2074)
	(2460,2057)(2467,2042)(2475,2029)
	(2486,2016)(2493,2008)(2501,2000)
	(2510,1993)(2521,1985)(2532,1977)
	(2545,1970)(2559,1963)(2575,1956)
	(2592,1950)(2611,1944)(2631,1939)
	(2652,1935)(2675,1931)(2699,1928)
	(2724,1926)(2750,1925)(2777,1925)
	(2805,1926)(2834,1928)(2865,1930)
	(2888,1933)(2912,1936)(2937,1940)
	(2962,1944)(2990,1948)(3018,1953)
	(3047,1959)(3077,1965)(3107,1971)
	(3139,1978)(3171,1985)(3204,1992)
	(3237,2000)(3270,2007)(3303,2015)
	(3336,2023)(3368,2031)(3399,2039)
	(3430,2046)(3460,2054)(3490,2061)
	(3518,2068)(3544,2075)(3570,2081)
	(3595,2087)(3618,2093)(3640,2098)
	(3661,2103)(3692,2110)(3721,2116)
	(3748,2122)(3774,2127)(3798,2131)
	(3821,2134)(3843,2137)(3862,2139)
	(3881,2140)(3898,2141)(3913,2141)
	(3927,2140)(3939,2139)(3951,2137)
	(3961,2135)(3970,2132)(3979,2129)
	(3987,2126)(4000,2120)(4012,2113)
	(4025,2104)(4038,2094)(4054,2081)
	(4071,2067)(4088,2052)(4105,2037)
	(4118,2024)(4127,2016)(4130,2013)(4131,2012)
\path(3487,2629)(3488,2628)(3492,2625)
	(3501,2616)(3514,2603)(3531,2587)
	(3548,2570)(3563,2553)(3577,2539)
	(3588,2526)(3597,2514)(3605,2504)
	(3611,2493)(3615,2486)(3619,2478)
	(3621,2470)(3624,2462)(3625,2453)
	(3625,2444)(3625,2434)(3623,2424)
	(3620,2413)(3616,2402)(3610,2391)
	(3604,2380)(3596,2368)(3587,2356)
	(3577,2344)(3566,2332)(3556,2321)
	(3545,2310)(3532,2298)(3518,2285)
	(3501,2271)(3483,2255)(3462,2238)
	(3439,2219)(3414,2199)(3387,2177)
	(3358,2154)(3330,2131)(3302,2109)
	(3276,2089)(3254,2072)(3237,2059)
	(3226,2050)(3219,2044)(3216,2042)
\path(3036,1899)(3033,1898)(3028,1895)
	(3018,1889)(3002,1881)(2981,1871)
	(2956,1857)(2926,1841)(2894,1824)
	(2860,1806)(2825,1787)(2792,1768)
	(2759,1750)(2728,1733)(2700,1716)
	(2673,1701)(2649,1686)(2628,1672)
	(2608,1659)(2591,1647)(2575,1635)
	(2560,1623)(2547,1612)(2535,1600)
	(2518,1583)(2504,1566)(2492,1548)
	(2481,1529)(2471,1508)(2463,1486)
	(2456,1461)(2449,1434)(2443,1407)
	(2439,1380)(2435,1355)(2432,1335)
	(2430,1320)(2429,1311)(2429,1308)(2429,1307)
\put(8745.500,8545.500){\arc{530.330}{4.5705}{6.4251}}
\put(8981.754,10776.099){\arc{3974.094}{1.7090}{6.1789}}
\put(9270.500,9370.500){\arc{530.330}{1.4289}{3.2835}}
\put(9116.750,10771.125){\arc{3348.170}{0.8369}{1.4563}}
\put(9179.575,10958.856){\arc{3565.745}{0.0707}{0.8260}}
\thinlines
\put(9008,10908){\ellipse{4800}{4800}}
\put(9008,10908){\ellipse{3000}{3000}}
\thicklines
\path(10508,10908)(11408,10908)
\path(10058,9858)(10733,9183)
\put(2108,7308){\makebox(0,0)[lb]{\smash{{{\SetFigFont{7}{8.4}{\rmdefault}{\mddefault}{\updefault}$\pi(1)$}}}}}
\put(8708,7308){\makebox(0,0)[lb]{\smash{{{\SetFigFont{7}{8.4}{\rmdefault}{\mddefault}{\updefault}$c_0=\pi(g_0)$}}}}}
\put(8708,108){\makebox(0,0)[lb]{\smash{{{\SetFigFont{7}{8.4}{\rmdefault}{\mddefault}{\updefault}$\pi(g_2^{-1})$}}}}}
\put(2108,108){\makebox(0,0)[lb]{\smash{{{\SetFigFont{7}{8.4}{\rmdefault}{\mddefault}{\updefault}$\pi(g_1)$}}}}}
\end{picture}
}%
\caption{\label{Bbraid1r5} Elements of the 3 string braid group on the cylinder.}\end{figure}

\begin{figure}\setlength{\unitlength}{0.00025000in}
\begingroup\makeatletter\ifx\SetFigFont\undefined%
\gdef\SetFigFont#1#2#3#4#5{%
  \reset@font\fontsize{#1}{#2pt}%
  \fontfamily{#3}\fontseries{#4}\fontshape{#5}%
  \selectfont}%
\fi\endgroup%
{\renewcommand{\dashlinestretch}{30}
\begin{picture}(17083,7795)(0,-10)
\thicklines
\put(3463.420,66.867){\arc{859.443}{4.5831}{6.4042}}
\put(4080.484,3685.920){\arc{5434.774}{0.8136}{1.4641}}
\put(4300.258,1394.258){\arc{832.287}{1.4024}{3.3100}}
\put(3847.658,3673.038){\arc{6420.574}{1.7082}{6.1774}}
\put(4276.604,3946.329){\arc{5538.033}{0.0637}{0.8677}}
\thinlines
\put(3911,3869){\ellipse{3014}{3014}}
\put(3904,3876){\ellipse{4800}{4800}}
\put(3890,3890){\ellipse{7764}{7766}}
\thicklines
\path(5404,3876)(6304,3876)
\path(3911,2369)(3911,2368)(3911,2363)
	(3912,2352)(3914,2335)(3916,2313)
	(3920,2291)(3924,2270)(3929,2251)
	(3935,2234)(3942,2219)(3950,2206)
	(3961,2193)(3968,2185)(3976,2177)
	(3985,2170)(3996,2162)(4007,2154)
	(4020,2147)(4034,2140)(4050,2133)
	(4067,2127)(4086,2121)(4106,2116)
	(4127,2112)(4150,2108)(4174,2105)
	(4199,2103)(4225,2102)(4252,2102)
	(4280,2103)(4309,2105)(4340,2107)
	(4363,2110)(4387,2113)(4412,2117)
	(4437,2121)(4465,2125)(4493,2130)
	(4522,2136)(4552,2142)(4582,2148)
	(4614,2155)(4646,2162)(4679,2169)
	(4712,2177)(4745,2184)(4778,2192)
	(4811,2200)(4843,2208)(4874,2216)
	(4905,2223)(4935,2231)(4965,2238)
	(4993,2245)(5019,2252)(5045,2258)
	(5070,2264)(5093,2270)(5115,2275)
	(5136,2280)(5167,2287)(5196,2293)
	(5223,2299)(5249,2304)(5273,2308)
	(5296,2311)(5318,2314)(5337,2316)
	(5356,2317)(5373,2318)(5388,2318)
	(5402,2317)(5414,2316)(5426,2314)
	(5436,2312)(5445,2309)(5454,2306)
	(5462,2303)(5475,2297)(5487,2290)
	(5500,2281)(5513,2271)(5529,2258)
	(5546,2244)(5563,2229)(5580,2214)
	(5593,2201)(5602,2193)(5605,2190)(5606,2189)
\path(4991,2796)(4992,2795)(4995,2792)
	(5002,2784)(5013,2772)(5026,2756)
	(5040,2741)(5053,2725)(5063,2712)
	(5073,2700)(5080,2689)(5086,2679)
	(5091,2669)(5094,2661)(5097,2654)
	(5099,2646)(5101,2638)(5102,2629)
	(5102,2620)(5101,2611)(5099,2600)
	(5095,2590)(5091,2579)(5086,2568)
	(5079,2557)(5071,2545)(5062,2533)
	(5052,2521)(5041,2509)(5031,2498)
	(5020,2487)(5007,2475)(4993,2462)
	(4976,2448)(4958,2432)(4937,2415)
	(4914,2396)(4889,2376)(4862,2354)
	(4833,2331)(4805,2308)(4777,2286)
	(4751,2266)(4729,2249)(4712,2236)
	(4701,2227)(4694,2221)(4691,2219)
\path(4511,2076)(4508,2075)(4503,2072)
	(4493,2066)(4477,2058)(4456,2048)
	(4431,2034)(4401,2018)(4369,2001)
	(4335,1983)(4300,1964)(4267,1945)
	(4234,1927)(4203,1910)(4175,1893)
	(4148,1878)(4124,1863)(4103,1849)
	(4083,1836)(4066,1824)(4050,1812)
	(4035,1800)(4022,1789)(4010,1777)
	(3993,1760)(3979,1743)(3967,1725)
	(3956,1706)(3946,1685)(3938,1663)
	(3931,1638)(3924,1611)(3918,1584)
	(3914,1557)(3910,1532)(3907,1512)
	(3905,1497)(3904,1488)(3904,1485)(3904,1484)
\path(6308,3890)(7761,3890)
\path(5583,2193)(6680,1100)
\put(12763.420,66.867){\arc{859.443}{4.5831}{6.4042}}
\put(13147.658,3673.038){\arc{6420.574}{1.7082}{6.1774}}
\put(14572.384,3695.600){\arc{3538.363}{6.2411}{7.4689}}
\thinlines
\put(13211,3869){\ellipse{3014}{3014}}
\put(13204,3876){\ellipse{4800}{4800}}
\put(13190,3890){\ellipse{7764}{7766}}
\thicklines
\path(13212,2356)(13212,1456)
\path(14262,2806)(14937,2131)
\path(14704,3876)(15604,3876)
\path(15608,3890)(17061,3890)
\path(13212,1456)(13212,1455)(13212,1451)
	(13213,1442)(13214,1426)(13215,1405)
	(13218,1381)(13221,1356)(13224,1332)
	(13228,1310)(13233,1289)(13238,1270)
	(13245,1253)(13253,1236)(13262,1218)
	(13268,1208)(13275,1198)(13282,1188)
	(13290,1177)(13300,1167)(13310,1156)
	(13322,1146)(13336,1135)(13351,1125)
	(13367,1115)(13386,1106)(13406,1097)
	(13428,1089)(13451,1082)(13477,1075)
	(13504,1069)(13532,1065)(13562,1061)
	(13594,1058)(13628,1057)(13663,1056)
	(13700,1057)(13738,1058)(13779,1061)
	(13808,1064)(13838,1067)(13869,1070)
	(13901,1075)(13934,1079)(13969,1084)
	(14005,1090)(14042,1096)(14081,1103)
	(14120,1110)(14160,1117)(14202,1125)
	(14244,1133)(14287,1141)(14331,1150)
	(14375,1159)(14419,1168)(14463,1177)
	(14508,1187)(14552,1196)(14596,1206)
	(14639,1215)(14682,1224)(14724,1234)
	(14766,1243)(14806,1252)(14845,1260)
	(14883,1269)(14920,1277)(14956,1285)
	(14990,1292)(15023,1299)(15055,1306)
	(15086,1313)(15115,1319)(15144,1324)
	(15188,1333)(15229,1341)(15268,1348)
	(15305,1354)(15340,1359)(15373,1363)
	(15405,1367)(15434,1369)(15462,1371)
	(15488,1372)(15513,1372)(15535,1372)
	(15556,1371)(15575,1368)(15592,1366)
	(15608,1362)(15622,1358)(15635,1354)
	(15648,1349)(15659,1344)(15670,1338)
	(15681,1332)(15694,1324)(15707,1315)
	(15720,1306)(15734,1295)(15749,1282)
	(15765,1269)(15782,1253)(15801,1236)
	(15821,1217)(15841,1197)(15861,1178)
	(15880,1159)(15896,1142)(15909,1129)
	(15918,1120)(15924,1115)(15926,1112)
\path(14937,2131)(14938,2130)(14941,2127)
	(14949,2120)(14961,2111)(14974,2100)
	(14987,2091)(15000,2083)(15012,2077)
	(15024,2072)(15037,2068)(15047,2066)
	(15058,2064)(15071,2063)(15086,2061)
	(15103,2060)(15122,2059)(15143,2058)
	(15166,2057)(15188,2057)(15208,2056)
	(15223,2056)(15232,2056)(15236,2056)(15237,2056)
\put(8412,3931){\makebox(0,0)[lb]{\smash{{{\SetFigFont{7}{8.4}{\rmdefault}{\mddefault}{\updefault}=}}}}}
\end{picture}
}%
\caption{\label{Bbraid6r1} Example of composition $g_1 g_0$, and ambient
  isotopy/ Reidemeister move on the product in a cylinder braid group.}\end{figure}
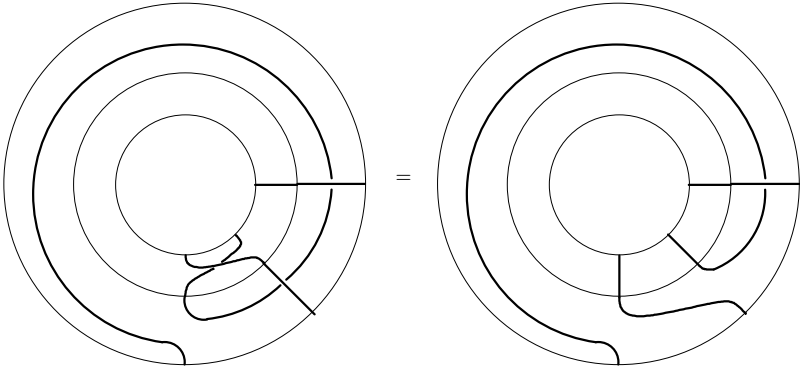

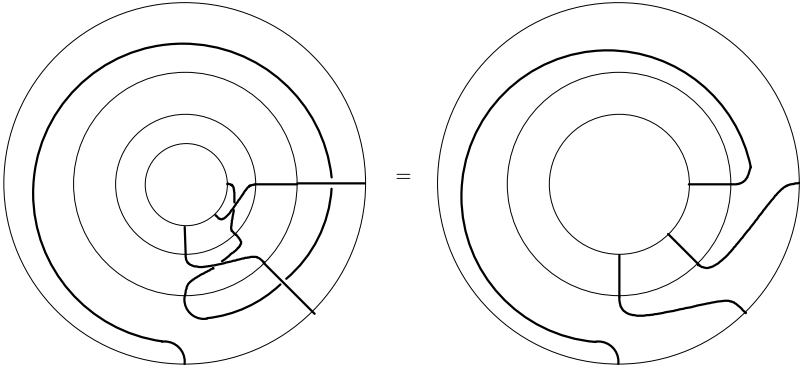
\begin{figure}\setlength{\unitlength}{0.00025000in}
\begingroup\makeatletter\ifx\SetFigFont\undefined%
\gdef\SetFigFont#1#2#3#4#5{%
  \reset@font\fontsize{#1}{#2pt}%
  \fontfamily{#3}\fontseries{#4}\fontshape{#5}%
  \selectfont}%
\fi\endgroup%
{\renewcommand{\dashlinestretch}{30}
\begin{picture}(17088,7795)(0,-10)
\thicklines
\put(3463.420,66.867){\arc{859.443}{4.5831}{6.4042}}
\put(4080.484,3685.920){\arc{5434.774}{0.8136}{1.4641}}
\put(4300.258,1394.258){\arc{832.287}{1.4024}{3.3100}}
\put(3847.658,3673.038){\arc{6420.574}{1.7082}{6.1774}}
\put(4276.604,3946.329){\arc{5538.033}{0.0637}{0.8677}}
\thinlines
\put(3911,3869){\ellipse{3014}{3014}}
\put(3904,3876){\ellipse{4800}{4800}}
\put(3890,3890){\ellipse{7764}{7766}}
\thicklines
\path(5404,3876)(6304,3876)
\path(3911,2369)(3911,2368)(3911,2363)
	(3912,2352)(3914,2335)(3916,2313)
	(3920,2291)(3924,2270)(3929,2251)
	(3935,2234)(3942,2219)(3950,2206)
	(3961,2193)(3968,2185)(3976,2177)
	(3985,2170)(3996,2162)(4007,2154)
	(4020,2147)(4034,2140)(4050,2133)
	(4067,2127)(4086,2121)(4106,2116)
	(4127,2112)(4150,2108)(4174,2105)
	(4199,2103)(4225,2102)(4252,2102)
	(4280,2103)(4309,2105)(4340,2107)
	(4363,2110)(4387,2113)(4412,2117)
	(4437,2121)(4465,2125)(4493,2130)
	(4522,2136)(4552,2142)(4582,2148)
	(4614,2155)(4646,2162)(4679,2169)
	(4712,2177)(4745,2184)(4778,2192)
	(4811,2200)(4843,2208)(4874,2216)
	(4905,2223)(4935,2231)(4965,2238)
	(4993,2245)(5019,2252)(5045,2258)
	(5070,2264)(5093,2270)(5115,2275)
	(5136,2280)(5167,2287)(5196,2293)
	(5223,2299)(5249,2304)(5273,2308)
	(5296,2311)(5318,2314)(5337,2316)
	(5356,2317)(5373,2318)(5388,2318)
	(5402,2317)(5414,2316)(5426,2314)
	(5436,2312)(5445,2309)(5454,2306)
	(5462,2303)(5475,2297)(5487,2290)
	(5500,2281)(5513,2271)(5529,2258)
	(5546,2244)(5563,2229)(5580,2214)
	(5593,2201)(5602,2193)(5605,2190)(5606,2189)
\path(4991,2796)(4992,2795)(4995,2792)
	(5002,2784)(5013,2772)(5026,2756)
	(5040,2741)(5053,2725)(5063,2712)
	(5073,2700)(5080,2689)(5086,2679)
	(5091,2669)(5094,2661)(5097,2654)
	(5099,2646)(5101,2638)(5102,2629)
	(5102,2620)(5101,2611)(5099,2600)
	(5095,2590)(5091,2579)(5086,2568)
	(5079,2557)(5071,2545)(5062,2533)
	(5052,2521)(5041,2509)(5031,2498)
	(5020,2487)(5007,2475)(4993,2462)
	(4976,2448)(4958,2432)(4937,2415)
	(4914,2396)(4889,2376)(4862,2354)
	(4833,2331)(4805,2308)(4777,2286)
	(4751,2266)(4729,2249)(4712,2236)
	(4701,2227)(4694,2221)(4691,2219)
\path(4511,2076)(4508,2075)(4503,2072)
	(4493,2066)(4477,2058)(4456,2048)
	(4431,2034)(4401,2018)(4369,2001)
	(4335,1983)(4300,1964)(4267,1945)
	(4234,1927)(4203,1910)(4175,1893)
	(4148,1878)(4124,1863)(4103,1849)
	(4083,1836)(4066,1824)(4050,1812)
	(4035,1800)(4022,1789)(4010,1777)
	(3993,1760)(3979,1743)(3967,1725)
	(3956,1706)(3946,1685)(3938,1663)
	(3931,1638)(3924,1611)(3918,1584)
	(3914,1557)(3910,1532)(3907,1512)
	(3905,1497)(3904,1488)(3904,1485)(3904,1484)
\path(6308,3890)(7761,3890)
\path(5583,2193)(6680,1100)
\put(12763.420,66.867){\arc{859.443}{4.5831}{6.4042}}
\put(12956.309,3608.806){\arc{6274.423}{1.6523}{6.0832}}
\thinlines
\put(13211,3869){\ellipse{3014}{3014}}
\put(13190,3890){\ellipse{7764}{7766}}
\put(3926,3857){\ellipse{1768}{1768}}
\put(13204,3876){\ellipse{4800}{4800}}
\thicklines
\path(13212,2356)(13212,1456)
\path(14262,2806)(14937,2131)
\path(3896,2957)(3911,2372)
\path(14704,3876)(15604,3876)
\path(13212,1456)(13212,1455)(13212,1451)
	(13213,1442)(13214,1426)(13215,1405)
	(13218,1381)(13221,1356)(13224,1332)
	(13228,1310)(13233,1289)(13238,1270)
	(13245,1253)(13253,1236)(13262,1218)
	(13268,1208)(13275,1198)(13282,1188)
	(13290,1177)(13300,1167)(13310,1156)
	(13322,1146)(13336,1135)(13351,1125)
	(13367,1115)(13386,1106)(13406,1097)
	(13428,1089)(13451,1082)(13477,1075)
	(13504,1069)(13532,1065)(13562,1061)
	(13594,1058)(13628,1057)(13663,1056)
	(13700,1057)(13738,1058)(13779,1061)
	(13808,1064)(13838,1067)(13869,1070)
	(13901,1075)(13934,1079)(13969,1084)
	(14005,1090)(14042,1096)(14081,1103)
	(14120,1110)(14160,1117)(14202,1125)
	(14244,1133)(14287,1141)(14331,1150)
	(14375,1159)(14419,1168)(14463,1177)
	(14508,1187)(14552,1196)(14596,1206)
	(14639,1215)(14682,1224)(14724,1234)
	(14766,1243)(14806,1252)(14845,1260)
	(14883,1269)(14920,1277)(14956,1285)
	(14990,1292)(15023,1299)(15055,1306)
	(15086,1313)(15115,1319)(15144,1324)
	(15188,1333)(15229,1341)(15268,1348)
	(15305,1354)(15340,1359)(15373,1363)
	(15405,1367)(15434,1369)(15462,1371)
	(15488,1372)(15513,1372)(15535,1372)
	(15556,1371)(15575,1368)(15592,1366)
	(15608,1362)(15622,1358)(15635,1354)
	(15648,1349)(15659,1344)(15670,1338)
	(15681,1332)(15694,1324)(15707,1315)
	(15720,1306)(15734,1295)(15749,1282)
	(15765,1269)(15782,1253)(15801,1236)
	(15821,1217)(15841,1197)(15861,1178)
	(15880,1159)(15896,1142)(15909,1129)
	(15918,1120)(15924,1115)(15926,1112)
\path(4556,3212)(4557,3211)(4561,3207)
	(4570,3197)(4583,3184)(4597,3170)
	(4611,3158)(4623,3147)(4634,3139)
	(4645,3134)(4656,3129)(4663,3128)
	(4670,3126)(4678,3126)(4687,3126)
	(4696,3127)(4706,3130)(4717,3134)
	(4729,3139)(4741,3146)(4754,3155)
	(4767,3165)(4781,3177)(4795,3190)
	(4810,3205)(4825,3221)(4841,3239)
	(4853,3254)(4866,3270)(4879,3287)
	(4893,3305)(4908,3325)(4923,3345)
	(4938,3367)(4955,3390)(4971,3413)
	(4988,3437)(5004,3461)(5021,3485)
	(5037,3508)(5053,3532)(5069,3555)
	(5084,3577)(5098,3598)(5112,3618)
	(5124,3637)(5137,3655)(5148,3671)
	(5159,3687)(5174,3709)(5188,3730)
	(5202,3748)(5214,3765)(5226,3780)
	(5237,3793)(5247,3805)(5256,3815)
	(5264,3823)(5272,3830)(5279,3836)
	(5286,3840)(5292,3844)(5299,3847)
	(5308,3851)(5318,3854)(5330,3857)
	(5344,3861)(5360,3864)(5377,3867)
	(5393,3869)(5405,3871)(5410,3872)(5411,3872)
\path(4811,3887)(4812,3887)(4818,3885)
	(4830,3883)(4844,3880)(4857,3876)
	(4868,3873)(4877,3870)(4886,3867)
	(4892,3865)(4898,3862)(4904,3858)
	(4910,3854)(4916,3848)(4922,3842)
	(4927,3834)(4933,3825)(4938,3815)
	(4942,3804)(4945,3791)(4949,3777)
	(4951,3765)(4952,3752)(4954,3737)
	(4955,3720)(4956,3700)(4957,3677)
	(4958,3652)(4959,3624)(4959,3594)
	(4960,3564)(4960,3537)(4961,3514)
	(4961,3497)(4961,3487)(4961,3483)(4961,3482)
\path(4991,2792)(4990,2793)(4986,2797)
	(4976,2805)(4963,2818)(4948,2831)
	(4935,2843)(4924,2854)(4916,2864)
	(4909,2872)(4904,2879)(4899,2887)
	(4896,2895)(4893,2904)(4891,2914)
	(4890,2926)(4890,2939)(4890,2953)
	(4892,2969)(4894,2986)(4896,3004)
	(4898,3018)(4900,3032)(4902,3048)
	(4905,3066)(4909,3088)(4912,3112)
	(4917,3139)(4922,3168)(4927,3200)
	(4932,3231)(4937,3260)(4941,3284)
	(4943,3302)(4945,3312)(4946,3316)(4946,3317)
\path(14906,2147)(14907,2147)(14910,2144)
	(14919,2139)(14932,2131)(14948,2122)
	(14966,2112)(14984,2103)(15001,2096)
	(15018,2091)(15035,2086)(15052,2084)
	(15071,2082)(15082,2082)(15093,2081)
	(15106,2082)(15119,2083)(15134,2085)
	(15149,2089)(15166,2093)(15184,2099)
	(15203,2106)(15224,2115)(15245,2126)
	(15268,2138)(15292,2153)(15317,2169)
	(15343,2187)(15369,2208)(15397,2230)
	(15425,2254)(15454,2280)(15483,2308)
	(15514,2339)(15546,2372)(15566,2394)
	(15587,2416)(15608,2440)(15630,2465)
	(15653,2491)(15676,2518)(15700,2546)
	(15725,2576)(15750,2606)(15776,2637)
	(15802,2670)(15829,2703)(15857,2737)
	(15884,2772)(15912,2807)(15941,2842)
	(15969,2878)(15997,2914)(16025,2951)
	(16054,2987)(16081,3022)(16109,3058)
	(16136,3093)(16163,3127)(16189,3161)
	(16214,3194)(16239,3226)(16263,3257)
	(16286,3288)(16309,3317)(16331,3345)
	(16352,3372)(16372,3398)(16392,3423)
	(16410,3447)(16429,3469)(16457,3505)
	(16484,3538)(16509,3569)(16534,3598)
	(16557,3625)(16579,3651)(16601,3675)
	(16621,3697)(16641,3718)(16660,3736)
	(16677,3753)(16694,3769)(16710,3783)
	(16724,3795)(16738,3805)(16751,3815)
	(16763,3823)(16775,3829)(16785,3835)
	(16796,3841)(16806,3845)(16816,3849)
	(16832,3856)(16848,3861)(16866,3866)
	(16885,3871)(16906,3876)(16930,3880)
	(16956,3885)(16983,3890)(17009,3894)
	(17032,3897)(17050,3900)(17061,3901)
	(17065,3902)(17066,3902)
\path(15611,3872)(15612,3872)(15615,3872)
	(15624,3872)(15639,3873)(15657,3874)
	(15679,3875)(15701,3876)(15722,3877)
	(15741,3879)(15758,3881)(15773,3884)
	(15786,3887)(15799,3890)(15811,3894)
	(15825,3900)(15838,3907)(15852,3915)
	(15866,3924)(15879,3934)(15893,3945)
	(15905,3957)(15917,3970)(15928,3983)
	(15939,3997)(15948,4010)(15956,4024)
	(15963,4037)(15969,4051)(15975,4065)
	(15982,4082)(15988,4101)(15995,4122)
	(16002,4145)(16010,4170)(16016,4194)
	(16022,4216)(16027,4232)(16030,4242)
	(16031,4246)(16031,4247)
\put(8412,3931){\makebox(0,0)[lb]{\smash{{{\SetFigFont{7}{8.4}{\rmdefault}{\mddefault}{\updefault}=}}}}}
\end{picture}
}%
\caption{\label{Bbraidg2g1g0} Example of composition $g_2 g_1 c_0$, and ambient
  isotopy/ Reidemeister move on the product in a cylinder braid group,
  showing that this is a twist element $\twist$.}\end{figure}

Let $\braido{n}$ denote the Artin braid group on the cylinder (or
annulus --- the correspondence between the cylinder and annulus versions is
trivial, cf. \cite{MartinSaleur93,Jones94b}, 
and we will use them interchangeably).
Figure~\ref{Bbraid1r5} shows some
elements of $\braido{3}$
(together with an assertion, to be verified later, of their preimages in
$\Artin{B_3}$   under a certain group homomorphism).
Figure~\ref{Bbraid6r1} illustrates composition in the cylinder braid group,
and the Reidemeister move \cite[III\S1]{Reidemeister48} of type 2 in
this context (\cite{Reidemeister48} provides a summary of and link to
Reidemeister's original works).
There is an obvious inclusion
$\glue: \braid{n}  \hookrightarrow \braido{n}$
got by identifying the right and left edges of the frame.
There is an obvious surjective homomorphism
$\squash: \braido{n} \rightarrow \braid{n}$
got by arranging for all the string endpoints to be gathered
on one side of the cylinder and then
squashing the cylinder flat with this side on top.
\footnote{Most of the groups we consider here contain $\braid{m}$ as a subgroup
at least for some $m$. 
For example, if $A_{m-1}$ is a full subgraph of $G$ then $\Artin{G}
\supset \braid{m}$. 
Where it is unambiguous to do so we will refer
to the elements which lie in this subgroup by their  $\braid{m}$
names (thus $g_1$ and so on).
}
Figure~\ref{Bbraidg2g1g0} shows that
$\twist = g_{n-1}\ldots g_2 g_1 c_0$
is a useful {\em twist} element.

\prl(cylgen)
Each of the sets $S = \{ c_0^{\pm 1}, g_1^{\pm 1}, g_2^{\pm 1},... \}$
and
$S' = \{ \twist^{\pm 1}, g_1^{\pm 1}, g_2^{\pm 1},... \}$
generates $\braido{n}$.
\end{pr}
{\em Proof:}
These sets generate each other so it is enough to prove for $S$.
Let $w$ be an arbitrary cylinder braid.
We may assume that it is drawn with no string tangent parallel to the
top frame, and it has a finite number of crossings.
Either it has no crossings, in which case it can evidently be
generated by $\twist^{\pm 1}$
(figure \ref{Bbraidg2g1g0}), or there exist a pair of
strings adjacent at the top of the diagram which cross each other
before crossing any other. This crossing can be removed by multiplying
by an appropriate element of $S$ (or finite product thereof). Since
$w$ is finite, iterating this process produces a braid with no
crossings (and hence generated by $\twist$). Thus the inverse of $w$
is generated by $S$, and so is $w$. \Qed

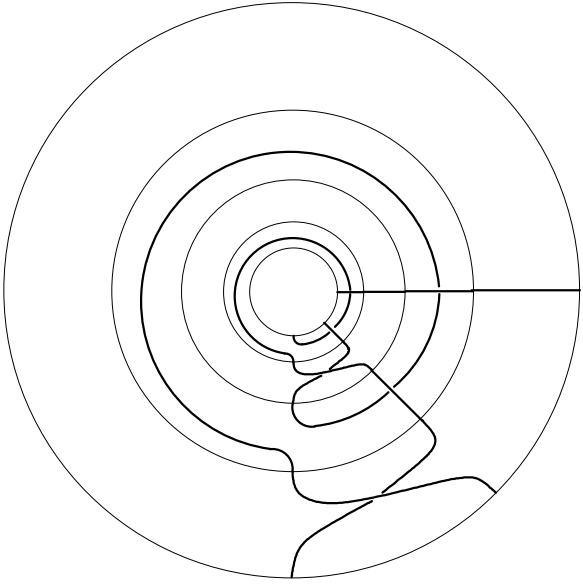
\begin{figure}\setlength{\unitlength}{0.00025000in}
\begingroup\makeatletter\ifx\SetFigFont\undefined%
\gdef\SetFigFont#1#2#3#4#5{%
  \reset@font\fontsize{#1}{#2pt}%
  \fontfamily{#3}\fontseries{#4}\fontshape{#5}%
  \selectfont}%
\fi\endgroup%
{\renewcommand{\dashlinestretch}{30}
\begin{picture}(12389,12389)(0,-10)
\thicklines
\put(6058.014,4665.124){\arc{336.388}{4.5871}{6.4031}}
\put(6384.162,5181.309){\arc{323.106}{1.4039}{3.3138}}
\put(6208.196,6066.906){\arc{2493.431}{1.7085}{6.1778}}
\put(5777.420,2347.867){\arc{859.443}{4.5831}{6.4042}}
\put(6394.484,5966.920){\arc{5434.774}{0.8136}{1.4641}}
\put(6614.258,3675.258){\arc{832.287}{1.4024}{3.3100}}
\put(6307.249,6042.093){\arc{2050.711}{0.8408}{1.4694}}
\put(6392.844,6167.079){\arc{2114.080}{0.0597}{0.8148}}
\put(6154.476,5951.852){\arc{6414.289}{1.7061}{6.1844}}
\put(6610.377,6212.339){\arc{5489.604}{0.0398}{0.8697}}
\thinlines
\put(6225,6150){\ellipse{3014}{3014}}
\put(6225,6150){\ellipse{1882}{1882}}
\put(6218,6157){\ellipse{4800}{4800}}
\put(6204,6171){\ellipse{7764}{7766}}
\put(6188,6187){\ellipse{12360}{12360}}
\thicklines
\path(7165,6150)(7729,6150)
\path(6882,5492)(7308,5067)
\path(7718,6157)(8618,6157)
\path(6225,4650)(6225,4649)(6225,4644)
	(6226,4633)(6228,4616)(6230,4594)
	(6234,4572)(6238,4551)(6243,4532)
	(6249,4515)(6256,4500)(6264,4487)
	(6275,4474)(6282,4466)(6290,4458)
	(6299,4451)(6310,4443)(6321,4435)
	(6334,4428)(6348,4421)(6364,4414)
	(6381,4408)(6400,4402)(6420,4397)
	(6441,4393)(6464,4389)(6488,4386)
	(6513,4384)(6539,4383)(6566,4383)
	(6594,4384)(6623,4386)(6654,4388)
	(6677,4391)(6701,4394)(6726,4398)
	(6751,4402)(6779,4406)(6807,4411)
	(6836,4417)(6866,4423)(6896,4429)
	(6928,4436)(6960,4443)(6993,4450)
	(7026,4458)(7059,4465)(7092,4473)
	(7125,4481)(7157,4489)(7188,4497)
	(7219,4504)(7249,4512)(7279,4519)
	(7307,4526)(7333,4533)(7359,4539)
	(7384,4545)(7407,4551)(7429,4556)
	(7450,4561)(7481,4568)(7510,4574)
	(7537,4580)(7563,4585)(7587,4589)
	(7610,4592)(7632,4595)(7651,4597)
	(7670,4598)(7687,4599)(7702,4599)
	(7716,4598)(7728,4597)(7740,4595)
	(7750,4593)(7759,4590)(7768,4587)
	(7776,4584)(7789,4578)(7801,4571)
	(7814,4562)(7827,4552)(7843,4539)
	(7860,4525)(7877,4510)(7894,4495)
	(7907,4482)(7916,4474)(7919,4471)(7920,4470)
\path(7305,5077)(7306,5076)(7309,5073)
	(7316,5065)(7327,5053)(7340,5037)
	(7354,5022)(7367,5006)(7377,4993)
	(7387,4981)(7394,4970)(7400,4960)
	(7405,4950)(7408,4942)(7411,4935)
	(7413,4927)(7415,4919)(7416,4910)
	(7416,4901)(7415,4892)(7413,4881)
	(7409,4871)(7405,4860)(7400,4849)
	(7393,4838)(7385,4826)(7376,4814)
	(7366,4802)(7355,4790)(7345,4779)
	(7334,4768)(7321,4756)(7307,4743)
	(7290,4729)(7272,4713)(7251,4696)
	(7228,4677)(7203,4657)(7176,4635)
	(7147,4612)(7119,4589)(7091,4567)
	(7065,4547)(7043,4530)(7026,4517)
	(7015,4508)(7008,4502)(7005,4500)
\path(6825,4357)(6822,4356)(6817,4353)
	(6807,4347)(6791,4339)(6770,4329)
	(6745,4315)(6715,4299)(6683,4282)
	(6649,4264)(6614,4245)(6581,4226)
	(6548,4208)(6517,4191)(6489,4174)
	(6462,4159)(6438,4144)(6417,4130)
	(6397,4117)(6380,4105)(6364,4093)
	(6349,4081)(6336,4070)(6324,4058)
	(6307,4041)(6293,4024)(6281,4006)
	(6270,3987)(6260,3966)(6252,3944)
	(6245,3919)(6238,3892)(6232,3865)
	(6228,3838)(6224,3813)(6221,3793)
	(6219,3778)(6218,3769)(6218,3766)(6218,3765)
\path(8622,6171)(10075,6171)
\path(7897,4474)(8994,3381)
\path(10048,6187)(12367,6187)
\path(6204,2304)(6204,2301)(6205,2293)
	(6206,2280)(6207,2260)(6209,2235)
	(6212,2205)(6216,2173)(6221,2139)
	(6226,2104)(6232,2071)(6239,2040)
	(6247,2011)(6255,1984)(6265,1958)
	(6275,1935)(6287,1913)(6301,1892)
	(6316,1872)(6333,1852)(6344,1839)
	(6356,1827)(6370,1815)(6384,1803)
	(6399,1790)(6415,1778)(6433,1766)
	(6451,1754)(6472,1742)(6493,1731)
	(6516,1719)(6541,1708)(6567,1698)
	(6595,1688)(6624,1678)(6654,1669)
	(6686,1661)(6720,1653)(6754,1646)
	(6790,1640)(6827,1634)(6866,1630)
	(6905,1626)(6945,1623)(6987,1621)
	(7030,1620)(7073,1620)(7118,1621)
	(7163,1623)(7210,1626)(7259,1630)
	(7309,1634)(7346,1639)(7384,1643)
	(7423,1648)(7464,1654)(7505,1660)
	(7548,1667)(7591,1674)(7636,1682)
	(7682,1690)(7729,1699)(7777,1708)
	(7825,1718)(7875,1728)(7926,1738)
	(7977,1749)(8029,1760)(8082,1772)
	(8135,1783)(8189,1796)(8243,1808)
	(8297,1820)(8351,1833)(8405,1846)
	(8459,1858)(8513,1871)(8566,1884)
	(8618,1897)(8670,1909)(8721,1922)
	(8771,1934)(8821,1946)(8869,1958)
	(8916,1970)(8963,1981)(9008,1992)
	(9052,2003)(9094,2014)(9135,2024)
	(9176,2034)(9215,2043)(9252,2052)
	(9289,2061)(9324,2069)(9359,2077)
	(9410,2089)(9460,2100)(9507,2110)
	(9553,2120)(9597,2129)(9639,2137)
	(9680,2144)(9720,2150)(9757,2156)
	(9794,2161)(9828,2165)(9861,2169)
	(9893,2171)(9923,2173)(9951,2174)
	(9978,2175)(10003,2174)(10026,2174)
	(10048,2172)(10069,2170)(10088,2167)
	(10106,2164)(10123,2160)(10139,2156)
	(10155,2151)(10169,2146)(10183,2140)
	(10197,2134)(10216,2126)(10235,2116)
	(10254,2105)(10273,2093)(10293,2079)
	(10314,2064)(10337,2047)(10361,2028)
	(10386,2006)(10413,1984)(10440,1960)
	(10467,1936)(10493,1913)(10516,1892)
	(10536,1874)(10551,1860)(10561,1850)
	(10567,1845)(10570,1842)
\path(8983,3406)(8985,3404)(8990,3398)
	(8999,3389)(9011,3375)(9027,3357)
	(9046,3335)(9067,3312)(9088,3287)
	(9110,3263)(9130,3239)(9148,3216)
	(9165,3195)(9181,3175)(9194,3157)
	(9206,3139)(9217,3123)(9226,3107)
	(9235,3092)(9242,3077)(9248,3064)
	(9253,3052)(9258,3039)(9262,3025)
	(9265,3012)(9267,2997)(9269,2982)
	(9269,2967)(9269,2951)(9267,2934)
	(9264,2917)(9260,2899)(9254,2881)
	(9247,2862)(9239,2843)(9229,2823)
	(9218,2803)(9205,2783)(9192,2762)
	(9176,2741)(9159,2720)(9141,2698)
	(9122,2675)(9100,2652)(9084,2635)
	(9067,2618)(9049,2600)(9029,2581)
	(9008,2561)(8985,2541)(8961,2519)
	(8935,2496)(8907,2471)(8877,2446)
	(8844,2418)(8810,2389)(8773,2359)
	(8734,2326)(8693,2293)(8649,2257)
	(8604,2221)(8558,2184)(8511,2146)
	(8464,2108)(8418,2071)(8372,2034)
	(8329,2000)(8289,1968)(8253,1940)
	(8221,1915)(8195,1893)(8173,1876)
	(8156,1863)(8144,1854)(8137,1848)
	(8133,1844)(8131,1843)
\path(7906,1663)(7904,1662)(7901,1660)
	(7894,1656)(7883,1650)(7867,1642)
	(7846,1631)(7820,1617)(7788,1599)
	(7751,1579)(7709,1557)(7663,1531)
	(7613,1504)(7559,1474)(7503,1444)
	(7446,1412)(7387,1379)(7328,1346)
	(7270,1314)(7213,1281)(7157,1249)
	(7102,1218)(7050,1188)(7000,1158)
	(6952,1130)(6907,1102)(6863,1076)
	(6823,1051)(6784,1026)(6748,1003)
	(6714,980)(6682,958)(6652,937)
	(6624,916)(6597,896)(6572,877)
	(6549,857)(6526,838)(6505,819)
	(6485,800)(6458,774)(6433,747)
	(6409,720)(6388,693)(6368,665)
	(6350,637)(6333,607)(6317,576)
	(6302,544)(6289,510)(6276,475)
	(6264,437)(6254,399)(6243,359)
	(6234,318)(6226,277)(6218,237)
	(6211,198)(6205,162)(6200,129)
	(6196,100)(6193,76)(6191,57)
	(6190,44)(6189,35)(6188,30)(6188,28)
\end{picture}
}%
\caption{\label{Bbraid5} Demonstration of the image of relation
  (\ref{4rel}) in a cylinder braid group.}\end{figure}

The interplay between $B$--type and periodic algebraic systems and boundary
conditions for YBE (cf. (\ref{4rel}),(\ref{3rel})) 
is neatly summed up by the following.
\prl(prop1)
There is a group homomorphism
\[
\pi : \Artin{B_{n+1}} \longrightarrow \braido{n}
\]
in which the images of the set $\{ g_0, g_1 , g_2, .. \}$ of 
generators are (the generators) as indicated in
figure~\ref{Bbraid1r5}.
\end{pr}
Figure~\ref{Bbraid5} verifies the special relation (\ref{4rel}) in this
realisation, in as much as it is manifest that the (outer) factor of
$\pi(g_1)$ commutes with the rest of the diagram.
Note from proposition \ref{cylgen} that $\pi$ is surjective.
(And see \cite{Brieskorn73,Allcock99}.)

It will be evident that there is a homomorphism from
$\Artin{\affine{A}_{n+1}}$
(with generator $\affine{g}_{n+1}$, say, where vertex $n+1$ is
adjacent to both 1 and $n$ in $\affine{A}_{n+1}$)
into $\braido{n+1}$.
This may be given in our $n=2$
example as $\affine{g}_3 \mapsto \twist g_1 \twist^{-1}$.

\subsection{On maps into the ordinary braid group}\label{->braidgp}

Recall that the pure braid group $\braid{n}'$ is normal in
$\braid{n}$, and that the quotient defines a surjection onto the
symmetric group $S_n$
$$ P: \braid{n} \rightarrow S_n . $$
For $p$ a partition of $\{1,2,..,n \}$, the subset of
permutations  which fix $p$ forms a subgroup, called the
{\em Young subgroup} $S_p$ of $S_n$.
We may extend this to define a subgroup $\braid{p}$
of $\braid{n}$ which fixes $p$ in the
sense that braid $b$ fixes $p$ if $P(b)$ does.

Note that for each $p_i$ a part of $p$
there is a natural `restricting' map from
$\braid{p}$ onto $\braid{|p_i|}$
which simply ignores all strings not in $p_i$.

For $m=1,2,...$ let $\braid{n+m}^m$ denote the subgroup of
$\braid{n+m}$ in which the first $m$ strings are pure.


Next we establish maps between $\Artin{B_n}$, $\braido{n}$ and
$\braid{n+1}^1$ which enable us to port information between them. This
is useful as each brings a particular utility to the problem of their
analysis ($\braido{n}$ has nice diagrams, and periodicity;
$\braid{n+1}^1$ forms a tower of subalgebras on varying $n$, and has
representations by restriction from $\braid{n}$; and  $\Artin{B_n}$
has the direct structural similarity with RE and the blob algebra (see
later)).

\begin{figure}\setlength{\unitlength}{0.00025000in}
\begingroup\makeatletter\ifx\SetFigFont\undefined%
\gdef\SetFigFont#1#2#3#4#5{%
  \reset@font\fontsize{#1}{#2pt}%
  \fontfamily{#3}\fontseries{#4}\fontshape{#5}%
  \selectfont}%
\fi\endgroup%
{\renewcommand{\dashlinestretch}{30}
\begin{picture}(11730,4837)(0,-10)
\thicklines
\put(9743.000,1665.000){\arc{2535.133}{2.2640}{3.5045}}
\thinlines
\texture{55888888 88555555 5522a222 a2555555 55888888 88555555 552a2a2a 2a555555 
	55888888 88555555 55a222a2 22555555 55888888 88555555 552a2a2a 2a555555 
	55888888 88555555 5522a222 a2555555 55888888 88555555 552a2a2a 2a555555 
	55888888 88555555 55a222a2 22555555 55888888 88555555 552a2a2a 2a555555 }
\shade\path(1808,1815)(233,240)(308,165)
	(1883,1740)(1808,1815)
\path(1808,1815)(233,240)(308,165)
	(1883,1740)(1808,1815)
\shade\path(8708,1815)(7133,240)(7208,165)
	(8783,1740)(8708,1815)
\path(8708,1815)(7133,240)(7208,165)
	(8783,1740)(8708,1815)
\thicklines
\put(2145.500,52.500){\arc{530.330}{4.5705}{6.4251}}
\put(2381.754,2283.099){\arc{3974.094}{1.7090}{6.1789}}
\put(2670.500,877.500){\arc{530.330}{1.4289}{3.2835}}
\put(2516.750,2278.125){\arc{3348.170}{0.8369}{1.4563}}
\put(2579.575,2465.856){\arc{3565.745}{0.0707}{0.8260}}
\thinlines
\put(2408,2415){\ellipse{4800}{4800}}
\put(2408,2415){\ellipse{3000}{3000}}
\thicklines
\path(3908,2415)(4808,2415)
\path(3458,1365)(4133,690)
\put(9045.500,52.500){\arc{530.330}{4.5705}{6.4251}}
\put(9045.500,927.500){\arc{525.595}{0.0476}{2.0132}}
\put(9301.488,1538.578){\arc{2516.567}{1.8062}{3.6175}}
\thinlines
\put(9308,2415){\ellipse{4800}{4800}}
\put(9308,2415){\ellipse{3000}{3000}}
\path(5408,2415)(6308,2415)
\path(6188.000,2385.000)(6308.000,2415.000)(6188.000,2445.000)
\thicklines
\path(10808,2415)(11708,2415)
\path(10358,1365)(11033,690)
\path(8183,2115)(8184,2117)(8187,2121)
	(8191,2128)(8199,2140)(8208,2155)
	(8221,2174)(8236,2197)(8253,2223)
	(8271,2252)(8291,2281)(8312,2312)
	(8333,2343)(8355,2373)(8377,2402)
	(8398,2431)(8420,2457)(8441,2482)
	(8462,2504)(8483,2524)(8503,2541)
	(8523,2554)(8541,2562)(8558,2565)
	(8575,2561)(8588,2549)(8597,2531)
	(8603,2508)(8606,2482)(8606,2452)
	(8605,2419)(8602,2385)(8598,2349)
	(8593,2312)(8587,2275)(8582,2240)
	(8576,2207)(8570,2179)(8566,2155)
	(8562,2137)(8560,2125)(8559,2118)(8558,2115)
\end{picture}
}%
\caption{\label{braid-in1} Inserting strings into the cylinder.}\end{figure}
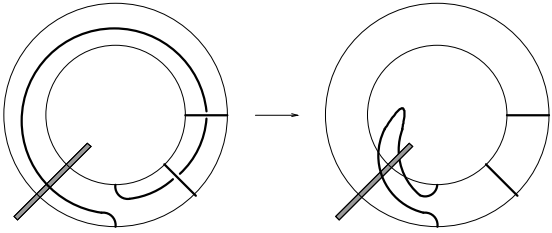

There is a mapping
$$ \squash_{\sm} :\braido{n} \rightarrow \braid{n+\sm}^{\sm} $$
like $\squash$, but which keeps track of which strings actually 
went round the back of the cylinder (i.e. it is injective).
Before squashing the cylinder completely flat we slide
an extra row of $\sm$ mutually non-crossing strings
into the hole, pushing them over so that they lie at,
say, the lefthand end of the row of strings in the squashed cylinder
(see figure \ref{braid-in1}).
For example $\squash_1 ( c_0 ) = g_1^2$,
$\squash_1 (g_i)=g_{i+1}$ ($i>0$).
To see that this map is injective note that the strings which went
round the back now go round the extra strings in the appropriate sense
(so the manoeuvre is reversible).
The image of this map is a nonempty subgroup of  $\braid{n+\sm}^{\sm}$ which
restricts, on the first $\sm$ strings, to the trivial group.
Note, then, that $\squash_1$ is an {\em isomorphism}.
We will again use these two realisations
interchangeably where no confusion arises.
(Cf. \cite{BaxterKellandWu76,BaxterTemperleyAshley78,MartinSaleur94a}.)
Indeed, for mapping the braid groups themselves the generalisation to
$\sm>1$ is effectively spurious.
We include it because we will later want to study the maps
induced by $\sigma_{\sm}$ on quotient algebras, and these maps do depend
on $\sm$ (and even on variations like 
attaching an idempotent to the first $\sm$ strings \cite{MartinWoodcock01pre}).



Next, consider the subgroup $J$ of $\braid{2n}$ consisting of
braids which are invariant under rotation about an axis passing
north to south, starting halfway
between the $n^{th}$ and $n+1^{th}$ northern endpoints, as in
figure~\ref{mirrorbraid1}.
\begin{figure}\setlength{\unitlength}{0.00041667in}
\begingroup\makeatletter\ifx\SetFigFont\undefined%
\gdef\SetFigFont#1#2#3#4#5{%
  \reset@font\fontsize{#1}{#2pt}%
  \fontfamily{#3}\fontseries{#4}\fontshape{#5}%
  \selectfont}%
\fi\endgroup%
{\renewcommand{\dashlinestretch}{30}
\begin{picture}(3474,4464)(0,-10)
\thicklines
\path(237,4137)(237,4134)(237,4128)
	(238,4118)(238,4102)(239,4080)
	(241,4053)(243,4022)(245,3987)
	(248,3950)(251,3913)(254,3876)
	(258,3839)(262,3804)(267,3771)
	(272,3739)(278,3710)(284,3682)
	(291,3656)(299,3631)(307,3607)
	(316,3583)(326,3560)(337,3537)
	(348,3516)(360,3495)(372,3473)
	(386,3451)(400,3429)(415,3406)
	(431,3383)(448,3359)(465,3335)
	(482,3311)(500,3287)(519,3262)
	(537,3237)(555,3212)(574,3187)
	(592,3163)(609,3139)(626,3115)
	(643,3091)(659,3068)(674,3045)
	(688,3023)(702,3001)(714,2979)
	(726,2958)(737,2937)(748,2914)
	(758,2891)(767,2867)(775,2843)
	(783,2818)(790,2792)(796,2764)
	(802,2735)(807,2703)(812,2670)
	(816,2635)(820,2598)(823,2561)
	(826,2524)(829,2487)(831,2452)
	(833,2421)(835,2394)(836,2372)
	(836,2356)(837,2346)(837,2340)(837,2337)
\path(237,2337)(237,2340)(237,2348)
	(237,2361)(238,2380)(238,2406)
	(239,2437)(240,2473)(241,2510)
	(242,2549)(243,2588)(245,2625)
	(247,2660)(249,2692)(251,2722)
	(253,2750)(256,2775)(259,2798)
	(262,2819)(266,2838)(270,2857)
	(275,2874)(281,2895)(287,2915)
	(295,2934)(304,2953)(314,2971)
	(326,2991)(339,3011)(354,3032)
	(370,3053)(387,3075)(405,3096)
	(421,3115)(435,3132)(447,3145)
	(455,3154)(460,3160)(462,3162)
\path(837,4137)(837,4134)(837,4126)
	(837,4113)(836,4094)(836,4068)
	(835,4037)(834,4001)(833,3964)
	(832,3925)(831,3886)(829,3849)
	(827,3814)(825,3782)(823,3752)
	(821,3724)(818,3699)(815,3676)
	(812,3655)(808,3636)(804,3617)
	(800,3599)(793,3576)(785,3554)
	(777,3533)(767,3512)(757,3493)
	(747,3474)(736,3455)(725,3438)
	(713,3423)(702,3408)(692,3395)
	(682,3383)(672,3373)(664,3364)
	(656,3356)(650,3349)(635,3335)
	(625,3325)(618,3318)(614,3314)(612,3312)
\path(1437,4137)(1437,4134)(1437,4128)
	(1438,4118)(1438,4102)(1439,4080)
	(1441,4053)(1443,4022)(1445,3987)
	(1448,3950)(1451,3913)(1454,3876)
	(1458,3839)(1462,3804)(1467,3771)
	(1472,3739)(1478,3710)(1484,3682)
	(1491,3656)(1499,3631)(1507,3607)
	(1516,3583)(1526,3560)(1537,3537)
	(1548,3516)(1560,3495)(1572,3473)
	(1586,3451)(1600,3429)(1615,3406)
	(1631,3383)(1648,3359)(1665,3335)
	(1682,3311)(1700,3287)(1719,3262)
	(1737,3237)(1755,3212)(1774,3187)
	(1792,3163)(1809,3139)(1826,3115)
	(1843,3091)(1859,3068)(1874,3045)
	(1888,3023)(1902,3001)(1914,2979)
	(1926,2958)(1937,2937)(1948,2914)
	(1958,2891)(1967,2867)(1975,2843)
	(1983,2818)(1990,2792)(1996,2764)
	(2002,2735)(2007,2703)(2012,2670)
	(2016,2635)(2020,2598)(2023,2561)
	(2026,2524)(2029,2487)(2031,2452)
	(2033,2421)(2035,2394)(2036,2372)
	(2036,2356)(2037,2346)(2037,2340)(2037,2337)
\path(1437,2337)(1437,2340)(1437,2348)
	(1437,2361)(1438,2380)(1438,2406)
	(1439,2437)(1440,2473)(1441,2510)
	(1442,2549)(1443,2588)(1445,2625)
	(1447,2660)(1449,2692)(1451,2722)
	(1453,2750)(1456,2775)(1459,2798)
	(1462,2819)(1466,2838)(1470,2857)
	(1475,2874)(1481,2895)(1487,2915)
	(1495,2934)(1504,2953)(1514,2971)
	(1526,2991)(1539,3011)(1554,3032)
	(1570,3053)(1587,3075)(1605,3096)
	(1621,3115)(1635,3132)(1647,3145)
	(1655,3154)(1660,3160)(1662,3162)
\path(2037,4137)(2037,4134)(2037,4126)
	(2037,4113)(2036,4094)(2036,4068)
	(2035,4037)(2034,4001)(2033,3964)
	(2032,3925)(2031,3886)(2029,3849)
	(2027,3814)(2025,3782)(2023,3752)
	(2021,3724)(2018,3699)(2015,3676)
	(2012,3655)(2008,3636)(2004,3617)
	(2000,3599)(1993,3576)(1985,3554)
	(1977,3533)(1967,3512)(1957,3493)
	(1947,3474)(1936,3455)(1925,3438)
	(1913,3423)(1902,3408)(1892,3395)
	(1882,3383)(1872,3373)(1864,3364)
	(1856,3356)(1850,3349)(1835,3335)
	(1825,3325)(1818,3318)(1814,3314)(1812,3312)
\path(2637,4137)(2637,4134)(2637,4128)
	(2638,4118)(2638,4102)(2639,4080)
	(2641,4053)(2643,4022)(2645,3987)
	(2648,3950)(2651,3913)(2654,3876)
	(2658,3839)(2662,3804)(2667,3771)
	(2672,3739)(2678,3710)(2684,3682)
	(2691,3656)(2699,3631)(2707,3607)
	(2716,3583)(2726,3560)(2737,3537)
	(2748,3516)(2760,3495)(2772,3473)
	(2786,3451)(2800,3429)(2815,3406)
	(2831,3383)(2848,3359)(2865,3335)
	(2882,3311)(2900,3287)(2919,3262)
	(2937,3237)(2955,3212)(2974,3187)
	(2992,3163)(3009,3139)(3026,3115)
	(3043,3091)(3059,3068)(3074,3045)
	(3088,3023)(3102,3001)(3114,2979)
	(3126,2958)(3137,2937)(3148,2914)
	(3158,2891)(3167,2867)(3175,2843)
	(3183,2818)(3190,2792)(3196,2764)
	(3202,2735)(3207,2703)(3212,2670)
	(3216,2635)(3220,2598)(3223,2561)
	(3226,2524)(3229,2487)(3231,2452)
	(3233,2421)(3235,2394)(3236,2372)
	(3236,2356)(3237,2346)(3237,2340)(3237,2337)
\path(2637,2337)(2637,2340)(2637,2348)
	(2637,2361)(2638,2380)(2638,2406)
	(2639,2437)(2640,2473)(2641,2510)
	(2642,2549)(2643,2588)(2645,2625)
	(2647,2660)(2649,2692)(2651,2722)
	(2653,2750)(2656,2775)(2659,2798)
	(2662,2819)(2666,2838)(2670,2857)
	(2675,2874)(2681,2895)(2687,2915)
	(2695,2934)(2704,2953)(2714,2971)
	(2726,2991)(2739,3011)(2754,3032)
	(2770,3053)(2787,3075)(2805,3096)
	(2821,3115)(2835,3132)(2847,3145)
	(2855,3154)(2860,3160)(2862,3162)
\path(3237,4137)(3237,4134)(3237,4126)
	(3237,4113)(3236,4094)(3236,4068)
	(3235,4037)(3234,4001)(3233,3964)
	(3232,3925)(3231,3886)(3229,3849)
	(3227,3814)(3225,3782)(3223,3752)
	(3221,3724)(3218,3699)(3215,3676)
	(3212,3655)(3208,3636)(3204,3617)
	(3200,3599)(3193,3576)(3185,3554)
	(3177,3533)(3167,3512)(3157,3493)
	(3147,3474)(3136,3455)(3125,3438)
	(3113,3423)(3102,3408)(3092,3395)
	(3082,3383)(3072,3373)(3064,3364)
	(3056,3356)(3050,3349)(3035,3335)
	(3025,3325)(3018,3318)(3014,3314)(3012,3312)
\path(837,2337)(837,2334)(837,2328)
	(838,2318)(838,2302)(839,2280)
	(841,2253)(843,2222)(845,2187)
	(848,2150)(851,2113)(854,2076)
	(858,2039)(862,2004)(867,1971)
	(872,1939)(878,1910)(884,1882)
	(891,1856)(899,1831)(907,1807)
	(916,1783)(926,1760)(937,1737)
	(948,1716)(960,1695)(972,1673)
	(986,1651)(1000,1629)(1015,1606)
	(1031,1583)(1048,1559)(1065,1535)
	(1082,1511)(1100,1487)(1119,1462)
	(1137,1437)(1155,1412)(1174,1387)
	(1192,1363)(1209,1339)(1226,1315)
	(1243,1291)(1259,1268)(1274,1245)
	(1288,1223)(1302,1201)(1314,1179)
	(1326,1158)(1337,1137)(1348,1114)
	(1358,1091)(1367,1067)(1375,1043)
	(1383,1018)(1390,992)(1396,964)
	(1402,935)(1407,903)(1412,870)
	(1416,835)(1420,798)(1423,761)
	(1426,724)(1429,687)(1431,652)
	(1433,621)(1435,594)(1436,572)
	(1436,556)(1437,546)(1437,540)(1437,537)
\path(837,537)(837,540)(837,548)
	(837,561)(838,580)(838,606)
	(839,637)(840,673)(841,710)
	(842,749)(843,788)(845,825)
	(847,860)(849,892)(851,922)
	(853,950)(856,975)(859,998)
	(862,1019)(866,1038)(870,1057)
	(875,1074)(881,1095)(887,1115)
	(895,1134)(904,1153)(914,1171)
	(926,1191)(939,1211)(954,1232)
	(970,1253)(987,1275)(1005,1296)
	(1021,1315)(1035,1332)(1047,1345)
	(1055,1354)(1060,1360)(1062,1362)
\path(1437,2337)(1437,2334)(1437,2326)
	(1437,2313)(1436,2294)(1436,2268)
	(1435,2237)(1434,2201)(1433,2164)
	(1432,2125)(1431,2086)(1429,2049)
	(1427,2014)(1425,1982)(1423,1952)
	(1421,1924)(1418,1899)(1415,1876)
	(1412,1855)(1408,1836)(1404,1817)
	(1400,1799)(1393,1776)(1385,1754)
	(1377,1733)(1367,1712)(1357,1693)
	(1347,1674)(1336,1655)(1325,1638)
	(1313,1623)(1302,1608)(1292,1595)
	(1282,1583)(1272,1573)(1264,1564)
	(1256,1556)(1250,1549)(1235,1535)
	(1225,1525)(1218,1518)(1214,1514)(1212,1512)
\path(2037,2337)(2037,2334)(2037,2328)
	(2038,2318)(2038,2302)(2039,2280)
	(2041,2253)(2043,2222)(2045,2187)
	(2048,2150)(2051,2113)(2054,2076)
	(2058,2039)(2062,2004)(2067,1971)
	(2072,1939)(2078,1910)(2084,1882)
	(2091,1856)(2099,1831)(2107,1807)
	(2116,1783)(2126,1760)(2137,1737)
	(2148,1716)(2160,1695)(2172,1673)
	(2186,1651)(2200,1629)(2215,1606)
	(2231,1583)(2248,1559)(2265,1535)
	(2282,1511)(2300,1487)(2319,1462)
	(2337,1437)(2355,1412)(2374,1387)
	(2392,1363)(2409,1339)(2426,1315)
	(2443,1291)(2459,1268)(2474,1245)
	(2488,1223)(2502,1201)(2514,1179)
	(2526,1158)(2537,1137)(2548,1114)
	(2558,1091)(2567,1067)(2575,1043)
	(2583,1018)(2590,992)(2596,964)
	(2602,935)(2607,903)(2612,870)
	(2616,835)(2620,798)(2623,761)
	(2626,724)(2629,687)(2631,652)
	(2633,621)(2635,594)(2636,572)
	(2636,556)(2637,546)(2637,540)(2637,537)
\path(2037,537)(2037,540)(2037,548)
	(2037,561)(2038,580)(2038,606)
	(2039,637)(2040,673)(2041,710)
	(2042,749)(2043,788)(2045,825)
	(2047,860)(2049,892)(2051,922)
	(2053,950)(2056,975)(2059,998)
	(2062,1019)(2066,1038)(2070,1057)
	(2075,1074)(2081,1095)(2087,1115)
	(2095,1134)(2104,1153)(2114,1171)
	(2126,1191)(2139,1211)(2154,1232)
	(2170,1253)(2187,1275)(2205,1296)
	(2221,1315)(2235,1332)(2247,1345)
	(2255,1354)(2260,1360)(2262,1362)
\path(2637,2337)(2637,2334)(2637,2326)
	(2637,2313)(2636,2294)(2636,2268)
	(2635,2237)(2634,2201)(2633,2164)
	(2632,2125)(2631,2086)(2629,2049)
	(2627,2014)(2625,1982)(2623,1952)
	(2621,1924)(2618,1899)(2615,1876)
	(2612,1855)(2608,1836)(2604,1817)
	(2600,1799)(2593,1776)(2585,1754)
	(2577,1733)(2567,1712)(2557,1693)
	(2547,1674)(2536,1655)(2525,1638)
	(2513,1623)(2502,1608)(2492,1595)
	(2482,1583)(2472,1573)(2464,1564)
	(2456,1556)(2450,1549)(2435,1535)
	(2425,1525)(2418,1518)(2414,1514)(2412,1512)
\path(237,2337)(237,537)
\path(3237,2337)(3237,537)
\thinlines
\dottedline{90}(12,4137)(3462,4137)(3462,537)
	(12,537)(12,4137)
\dashline{120.000}(1737,4437)(1737,12)
\end{picture}
}%
\caption{\label{mirrorbraid1} Element of $J \subset \braid{6}$, showing
  (dashed) symmetry axis.}\end{figure}
There is an injective homomorphism
$$\cable: \braid{n} \rightarrow J$$
\eql(gmirror)
\cable: g_i \mapsto g_{n-i}g_{n+i} . 
\eq
This extends to a homomorphism
$$\cable: \Artin{B_{n+1}}   
\rightarrow J$$
by
$$\cable: g_0 \mapsto g_n . $$
Physicists will recognise an analogy in this with the 
{\em method of images}. 
There is a similar extension of the cabling map \cite{MartinWoodcock01pre}.

Without the extension, the map $\cable$ is essentially the group
comultiplication $\Delta:\braid{n}\rightarrow\braid{n}\times\braid{n}$
embedded, Young subgroup style, in $\braid{2n}$.
Recall that this equips the group algebra with the property of
bialgebra (indeed Hopf algebra); and implies that the category of left
modules is closed under tensor products (see \cite{Joseph95} for
example).
There is a generalisation of this (see
\cite[\S A(iii)]{Martin89b}\cite{MartinWoodcock01pre}) which enables us to
close the sum over $q \in \C$ of categories of left $T_n(q)$--modules
under tensor products. It is possible to extend the representation
obtained by tensoring two copies of the ordinary spin--chain
representation (as in equation(\ref{rrrep})) to a representation of
$\braido{n}$ \cite{MartinWoodcock01pre}.
We will recall the precise construction in \S\ref{cablingX}. 
This is in particular 
a faithful 2--parameter representation of the blob algebra $b_n$ 
\cite{MartinSaleur94a}, which
is a quotient of $\braido{n}$ which explicitly solves RE 
--- see \S\ref{SSolutions}. 
As such this representation is arguably
the most interesting candidate for studying spin--chains with boundary currently
available. 
There are other possibilities, however, as  we now summarize. 


\subsection{Quotient algebras and representations}\label{ss2}
The above discussion gives us a number of recipes for constructing
representations of cylinder algebras from those of $\C\braid{n}$.
Many $\C\braid{n}$ representations may be used to construct exactly
solvable models, so applying the recipes to these should provide good
candidates for ESMs with more general boundary conditions.
Unfortunately these representations have important properties
which are not necessarily preserved by passage to the cylinder.
When $\C\braid{n}$ is used to solve the YBE it is never, physically, a
faithful representation which appears (and the vanishing of the
annihilator is {\em used} in the solution).
Indeed, on physical representations each $g_s$ has a finite spectrum.

If each $g_s$ has spectrum of order 2 then we are in the realm of {\em
  generic algebras} \cite{Humphreys90} (natural generalisations of the
corresponding Coxeter systems $(W,S)$ in which, of course, $g_s^2=1$
for all $s \in S$). In a generic algebra $g_s$ and $g_t$ have the same
spectrum if $s,t$ conjugate in $W$. Thus in the $A_n$ case each $g_s$
has the same spectrum --- we write
\eql(hecke rel)
(g_i - q)(g_i + q^{-1})=0
\eq
whereupon we have the ordinary Hecke algebra $H_n(q)$
\cite{KazhdanLusztig79}.
Although  $H_n(q)$ is a relatively tiny vestige of $\C\braid{n}$, even this
algebra is never faithfully represented in physical representations
(and no global limit of the whole of $H_n(q)$ is known).
A natural example of a quotient of $\C\braid{n}$ which {\em does} have
a global limit is the Temperley--Lieb algebra \cite{Martin91}.


We may assume that a similar situation pertains in the `affine' case. 
Applying (\ref{hecke rel}) to $\C\Artin{B_n}$ we get an {\em affine
  Hecke algebra} \cite{Cherednik91}, again too big to be physical. 
A number of potentially suitable quotients are discussed in
\cite{MartinWoodcockLevy00,MartinWoodcock01pre}.
The $N=2$ case (an affine equivalent of Temperley--Lieb) is the
aforementioned {\em blob algebra}. It has been examined
in some detail from the ordinary representation theory viewpoint
\cite{MartinSaleur94a}.
On the other hand, 
while $\C\braid{n}$ and its quotients all have a natural
inclusion via $A_n \subset A_{n+1}$, and a number of physically useful
representations are known,
embedding cylinder algebras in towers is
somewhat harder. The preceding discussion provides solutions to this
problem by building cylinder algebras out of ordinary ones. The price
paid is that while these constructions work at the level of braids,
they do not in general factor through the quotients which we are
obliged to restrict to physically.
The remainder of this paper is concerned with finding cases which
{\em  do} factor, and using these to solve the reflection equation. 
We typically have some variant of the following picture:
\[
\xymatrix{ \C\braido{n} \ar@{->}[d]^{\Psi^2} \ar@{->}[r]^{\sigma_.} \ar@{->}[rrd]
                             & \C\braid{n+\sm}^{\sm} \ar@{^{(}->}[r]  
                             & \C\braid{n+\sm} \ar@{->}[d] \\
 b_n  \ar@{.>}[rr]^{\Theta ?} 
                          &   & \End(V^{n+\sm}) . }
\]
Here $\sigma_.$ represents any of the maps constructed in \S\ref{->braidgp}; 
the diagonal map is defined by the commutativity of the upper
triangle; $\Psi^2$ is the quotient map to the blob algebra 
(see \S\ref{SSolutions}) or some other suitable quotient; 
and $\Theta$ is the representation of $b_n$ we get {\em if}
the diagonal map factors through $b_n$. 


Solutions which do not start with XXZ, or do not 
 end up in the blob quotient, raise rather different problems, 
and will be examined in a separate paper.

\section{The abstract blob algebra solution}\label{SSolutions}
\newcommand{\rr}{R_}
\newcommand{\U}{U_}
\newcommand{\ee}{e}
\newcommand{\Uone}{\U1}
\newcommand{\Uzero}{\ee}
\newcommand{\kap}{\kappa}
\newcommand{\kk}{K}
\newcommand{\T}{\theta_}
\newcommand{\tlambda}{\T}
\newcommand{\D}{\delta}
\newcommand{\sh}[1]{sh(#1)\;}
\newcommand{\ch}[1]{ch(#1)\;}
\newcommand{\Tr}{\mbox{Tr}}
\newcommand{\pvac}{\; |\;\ran}
\newcommand{\defsign}{-}
\newcommand{\minusdefsign}{+}
\newcommand{\yy}{m}
\newcommand{\yz}{y}
\newcommand{\sep}{\zeta}
\newcommand{\mT}{\!\theta_}
In this section we look for solutions to the reflection equation based on the
special representations of B--braids discussed above.
We show that the abstract blob algebra provides a meta--solution in the
same sense as the \TL\ algebra does for the ordinary YBE.

The blob algebra $b_n=b_n(q,\yy)$ may be defined by 
generators $\U1,\U2,...\U{n-1}$ and $\ee$, and relations:
\eql(r1) \U{i} \U{i} = \D \U{i} \eq
\eql(r2) \U{i} \U{i\pm 1} \U{i} = \U{i} \eq
\eql(r3) [\U{i},\U{j}]=0 \hspace{1in} |i-j|\neq 1 \eq
(so far we have the ordinary Temperley--Lieb algebra with $-\delta=q+q^{-1}$)
\eql(r4) \ee \ee = \D_{e} \ee \eq
\eql(r5) \U1 \ee \U1 = \kap \U1 \eq
$$ [\U{i} , \ee] = 0 \hspace{1in} i\neq 1 $$
Note that we are free to renormalize $\ee$, changing only $\D_{e}$ and
$\kap$ (by the same factor), thus from $\D,\D_{e},\kap$ there are
really only two relevant parameters. It will be natural later on to
reparameterize so that
they
are related (they only depend on $q$ and $\yy$), but it will be convenient to
treat them separately for the moment, and leave $\yy$ hidden.

Assuming for the moment that we have some viable representation of
this algebra we may proceed as follows.
Setting
$$ \rr{1}(\T{1} \pm \T2) = a_{\pm}\One + b_{\pm}\Uone $$
\be \kk(\T{i}) = x_i\One + y_i\Uzero  \label{ansatz1} \eeq
the reflection equation
$$ \rr{1}(\T1 -\T2 ) \kk(\T1 ) \rr{1}(\T1 +\T2 ) \kk(\T2 )
=  \kk(\T2 )  \rr{1}(\T1 +\T2 ) \kk(\T1 )  \rr{1}(\T1 -\T2 ) $$
becomes
$$ ( a_- \One + b_- \Uone ) (x_1 \One + y_1 \Uzero )
   ( a_+ \One + b_+ \Uone ) (x_2 \One + y_2 \Uzero )
$$ $$
=  (x_2 \One + y_2 \Uzero )  ( a_+ \One + b_+ \Uone )
   (x_1 \One + y_1 \Uzero ) ( a_- \One + b_- \Uone ) $$
and hence
$$  ( a_- x_1 \One + a_- y_1 \Uzero + b_- x_1 \Uone  + b_- y_1 \Uone \Uzero )
    ( a_+ x_2 \One + a_+ y_2 \Uzero + b_+ x_2 \Uone  + b_+ y_2 \Uone \Uzero )
$$ $$
=   ( a_+ x_2 \One + a_+ y_2 \Uzero + b_+ x_2 \Uone  + b_+ y_2 \Uzero \Uone )
    ( a_- x_1 \One + a_- y_1 \Uzero + b_- x_1 \Uone  + b_- y_1 \Uzero \Uone )
$$
and hence
\begin{eqnarray*}
\begin{array}{r}
a_- a_+ x_1 x_2 \; \One \\ +
a_- a_+ (x_1 y_2 +x_2 y_1 +\D_e y_1 y_2) \; \Uzero \\ +
(a_- b_+ +a_+ b_- +\D b_+ b_-) x_1 x_2 \; \Uone  \\ +
((a_- b_+ +a_+ b_- +\D b_+ b_-) x_1 y_2 + \hspace{.99cm} \\
          a_+ b_- (x_2 y_1 +\D_e y_1 y_2)) \; \Uone\Uzero \\ \\ +
a_- b_+ x_2 y_1 \; \Uzero\Uone \\ +
a_- b_+ y_1 y_2 \; \Uzero\Uone\Uzero \\ +
b_- b_+ y_1 x_2 \; \Uone\Uzero\Uone \\ +
b_- b_+ y_1 y_2 \; \Uone\Uzero\Uone\Uzero \\
\end{array}
\;\;\; = \;\;\; \begin{array}{r}
a_- a_+ x_1 x_2 \; \One \\ +
a_- a_+ (x_1 y_2 +x_2 y_1 +\D_e y_1 y_2) \; \Uzero \\ +
(a_- b_+ +a_+ b_- +\D b_+ b_-) x_1 x_2 \; \Uone  \\ \\ +
a_- b_+ x_2 y_1 \; \Uone\Uzero \\ +
((a_- b_+ +a_+ b_- +\D b_+ b_-) x_1 y_2 + \hspace{.99cm} \\
          a_+ b_- (x_2 y_1 +\D_e y_1 y_2)) \; \Uzero\Uone \\ +
a_- b_+ y_1 y_2 \; \Uzero\Uone\Uzero \\ +
b_- b_+ y_1 x_2 \; \Uone\Uzero\Uone \\  +
b_- b_+ y_1 y_2 \; \Uzero\Uone\Uzero\Uone
\end{array}
\end{eqnarray*}
Now applying relation(\ref{r5}) this becomes
$$ ((a_- b_+ +a_+ b_- +\D b_+ b_-) x_1 y_2 +
          a_+ b_- (x_2 y_1 +\D_e y_1 y_2)
          - a_- b_+ x_2 y_1  + \kap b_- b_+ y_1 y_2) \;
          [\Uzero,\Uone] =0 $$
Dividing by $y_1 y_2$ and putting $k_i = \frac{x_i}{y_i}$ we have
$$
(a_- b_+ + a_+ b_- + \D b_- b_+) k_1 + (-a_- b_+ + a_+ b_-) k_2
+(\D_e a_+ b_- + \kap b_+ b_-) =0
$$
so
$$ A_2 k_2 = A_1 k_1 + B $$
where $A_1 =(a_- b_+ + a_+ b_- + \D b_- b_+)$,
$B=(\D_e a_+ b_- + \kap b_+ b_-)$
and $A_2 =(a_- b_+ - a_+ b_-)$.
Since $k_i$ can depend only on $\T{i}$ this equation is required to
separate for a solution.


Recalling that  $q=e^{\mu i}$, then  $a_{\pm}=\sh{\mu(\T1\pm\T2+i)},
b_{\pm}=\sh{\mu(\T1\pm\T2)}$ are inherited from the global YB
solution. Thus 
$$ A_1 \mbox{\footnotesize $=\sh{\mu(\mT1-\mT2+\! i)\!}\sh{\mu(\mT1+\mT2)\!}
      +\!\sh{\mu(\mT1+\mT2+\! i)\!}\sh{\mu(\mT1-\mT2)\!}
      -\! 2\sh{\mu(\mT1+\mT2)\!}\sh{\mu(\mT1-\mT2)\!}\ch{\mu i} $} $$ $$
\hspace{.3in}    =\sh{\mu 2 \T1}\sh{\mu i} $$
$$ A_2 = \sh{\mu 2 \T2}\sh{\mu i} $$
$$ B = \D_e \sh{\mu(\T1-\T2)}\sh{\mu(\T1+\T2+i)}
      +\kap \sh{\mu(\T1-\T2)}\sh{\mu(\T1+\T2)} $$ $$
= \frac{1}{2} \left(
   \D_e (\ch{\mu(2\T1 +i)} - \ch{\mu(2\T2 +i)})
  +\kap (\ch{\mu(2\T1 )} - \ch{\mu(2\T2 )})
\right) $$
so we {\em may} separate to obtain
$$
\sh{\mu 2 \T{j}}\sh{\mu i} k_j
= \frac{-1}{2} (\D_e \ch{\mu(2\T{j} +i)}
               +\kap \ch{\mu(2\T{j} )} +\ch{\mu 2 i \sep})  $$
where $\sep$ is the (arbitrary) constant of separation.
At this point we have established a {\em solution to RE} (or rather a
meta--solution which produces a solution for each representation of
$b_n$). The blob algebra is a quotient of a special case of the
algebras shown to solve RE in
\cite{LevyMartin94,MartinWoodcockLevy00}, which guarantees
that it gives a solution in principle. However the precise form of
$b_n$ leads to a significant 
and crucial simplification in parameterization cf. the
general case. This is even more striking when we apply the
parameterization known from representation theory, as follows.

Recall $[\yy]=\frac{\sh{\yy \mu i}}{\sh{\mu i}}$.
In the abstract form a natural parameterisation of the two parameter
algebra $b_n$ is
$\D=-[2]$, $\D_e=-[\yy]$, $\kap = [\yy \defsign 1]$
(the two parameters are $q$ and $\yy$), and hence
$$
\sh{\mu 2 \T{j}}\sh{\mu i} k_j
= \frac{-1}{2} \left( \frac{ - \sh{\mu \yy i} \ch{\mu(2\T{j} +i)}
                             + \sh{\mu(\yy i \defsign i)} 
                                      \ch{\mu2\T{j}} }{\sh{\mu i}}
               + \ch{\mu 2 i \sep} \!\! \right)  $$
$$ =  \frac{1}{2} \left(  \ch{\mu(2\T{j} \minusdefsign \yy i)}
               - \ch{\mu 2 i \sep} \! \right)  $$
and hence
\eql(k-j)
k_j =  \frac{x_j}{y_j}
=  \frac{\sh{\mu(\T{j}+i(\frac{\minusdefsign \yy}{2}+\sep))}
                 \sh{\mu(\T{j}+i(\frac{\minusdefsign \yy}{2}-\sep))}}%
{\sh{\mu 2 \T{j}}\sh{\mu i}} .
\eq
Specifically we take  
\be x_j = x(\tlambda{j};\yy)
        = \sh{ \mu(\tlambda{j}+{i\yy \over 2} +i\sep )}
          \sh{ \mu(\tlambda{j}+{i\yy \over 2} -i\sep) } \,
\eeq \be 
\qquad \yz_j = z(\tlambda{j}) = \sh{\mu i} \sh{2\mu \tlambda{j}}
\,. \label{ansatz2}\eeq 
(We see that $\yy$ has the role of boundary parameter.)

Note that
$$
\kk(\T{}) \kk(-\T{})
\propto k(\T{}) k(-\T{}) 1 + (k(\T{}) + k(-\T{}) +\D_{e} ) \ee
=k(\T{}) k(-\T{}) 1 .
$$

\section{Realization via $\squash_{\sm}$ (auxiliary strings)}
It remains to construct representations suitable for forming the Bethe
ansatz. Our approach is to use the representations of the ordinary
Temperley--Lieb algebra for which there exists a Bethe ansatz
(we will concentrate on the XXZ representation) and pull
them through to the blob case using the tools in \S\ref{Sbraids}.
(Another approach would be to generalise \cite{MartinSaleur94c}, but
we do not consider that here.)
As noted in \S\ref{Sbraids} 
we have to check that this procedure preserves the appropriate
quotient inside $\C\Artin{B_n}$. 
In {\em general} it does not. 
The first cases we consider in which it {\em does} 
are the cases of $\squash_{\sm}$ in which $\sm=0,1$. 
The most obvious relation obeyed by $b_n$ cf. $\C\Artin{B_n}$ is
(\ref{r4}). The representation of $\ee$ will be a linear combination of
that of 1 and $c_0$, so we require the representation of $c_0$ to have at most two
eigenvalues. 
For $\squash_{\sm}$ (and general $q$) it is easy to check that
this holds for $\sm=0,1$ only. Case $\sm=0$ is the trivial solution ($K
\propto 1$, $\yy=1$), so we will focus on $\sm=1$.
The XXZ representation of $T_{\sites}(q)$ depends only on $q$, so the
representation pulled through $\squash_1$ also depends only on $q$,
thus $\yy$ must be fixed. Comparing (\ref{r4}), (\ref{r5}) and
$\XR(\squash_1(c_0))$ we see that $\yy=2$.

Using the XXZ representation for $T_{n+\sm}(q)$ as in eqn.(\ref{u}) 
we have that $\Theta: b_{n}\to T_{n+1} \to \End(V^{n+1})$ is given by 
$\Theta(\Uzero) = {\calR}(U_{1})$, 
$\Theta(U_{i}) = \calR(U_{i+1})$. 
Then using (\ref{ansatz1}),(\ref{ansatz2}) 
the $K$--matrix becomes
\be K(\lambda)=\left(
\begin{array}{cccc}
    x(\lambda;2)                                  \\
    &         w^{-}(\lambda)  & z(\lambda)                    \\
    &         z(\lambda) & w^{+}(\lambda)           \\
    &           &           & x(\lambda;2)

\end{array} \right) \,, \label{K1}
\eeq 
with $x(\lambda;2)$, $z(\lambda)$ given by (\ref{ansatz2}), and 
\be 
w^{\pm} (\lambda) &=& \sinh \mu(\lambda+i\sep)
                 \sinh \mu(\lambda -i\sep)+e^{\pm 2 \mu \lambda} \sinh^{2} (i\mu).
\label{ab} \eeq

The $K$--matrix  can be written in the following $ 2 \times 2$ form:
\be  K(\lambda) &=&\left(
\begin{array}{cc}
\underline{\alpha}(\lambda) &\underline{\beta}(\lambda)  \\
\underline{\gamma}(\lambda) & \underline{\delta}(\lambda)
\end{array} \right)
\non \\
&=& \left(
\begin{array}{cc}
x(\lambda;2)1-{1 \over 2} e^{i \mu } z(\lambda)(1- \sigma^{z}) 
                      & z(\lambda) \sigma^{-}  \\
z(\lambda) \sigma^{+} &x(\lambda;2)1-{1 \over 2} e^{-i \mu }
z(\lambda)(1+ \sigma^{z})
\end{array} \right) \label{K2} \eeq 
where $\sigma^{z}$, $\sigma^{\pm}$ act on a two dimensional space $V_{e}$. 
NB, This means that we extend the space
on which the transfer matrix acts from $2^{n}$--dimensional to $2^{n+1}$. 
This can be considered as a system with {\em enhanced space} 
(cf. \cite{BilsteinRittenberg00,BaseilhacDelius01}), 
i.e. it is as if we added an
extra site, with inhomogeneity $i\sep$, to the original spin chain.
The situation is similar in quantum impurity problems (see e.g. 
\cite{TsvelickWiegmann83,Fendley93,GeGouldLinksZhou99,FrahmSlavnov99,CastroFringGohmann02}). 


Suppose we are considering a system in which 
the underlying bulk model is a spin chain on $V^n$. 
Then a solution to RE is called `$\C$--number representation' if 
$K(\lambda)$ is an $\rank \times \rank$ matrix with complex entries 
\cite{\fk,GhoshalZamolodchikov94a,\DVGRi,deVegaGonzalezruiz94a,deVegaGonzalezruiz93}. 
More generally, it will be evident 
from figures~\ref{braidRE1},\ref{braidRE2},\ref{braidRE3} 
that given any $K(\lambda)$ which satisfies RE as in 
(\ref{boundaryYB0B}), the `factorized $K$--matrix' 
\be K_{f}(\lambda) = R(\lambda+i\sep) K(\lambda) R(\lambda -i\sep)
\label{kd} \eeq 
where $R$ is given by (\ref{xxz}),
is also a solution of RE. 
It is conjectured \cite{Kuznetsov01} that every
solution of RE is some iteration of this construction, with a 
$\C$--number representation as base. 
Our solution (\ref{K1}) is of this factorized form with $K=1$.

The eigenvalues of the corresponding open transfer matrix
(\ref{transfer10}) can be found via the algebraic Bethe ansatz method.

\subsection{The Bethe ansatz solution}

Here we show explicitly how the Bethe ansatz can be 
applied in the case of these `dynamical' 
\cite{FrahmSlavnov99} boundary conditions. 
(The analysis in this case is much closer to the usual setup than the
`cabled' case we will consider in \S\ref{cablingX}. 
We include it, since it also
serves the purpose of providing a preparatory review.)
We define the transfer matrix as in equation (\ref{transfer10}). 

The next step is to diagonalize the transfer matrix (\ref{transfer10})
using the algebraic Bethe ansatz method. The $\cal T$--matrix
(\ref{monodromyo}) has the form 
\be {\cal T}_{0}(\lambda) = \left( \begin{array}{cc}
 A(\lambda) & B'(\lambda)  \\
 C(\lambda) & D(\lambda)
\end{array} \right)
 \left( \begin{array}{cc}
\underline{\alpha}(\lambda) & \underline{\beta}(\lambda)  \\
\underline{\gamma}(\lambda) & \underline{\delta}(\lambda)
\end{array} \right)
 \left( \begin{array}{cc}
A(\lambda) & B(\lambda)  \\
C'(\lambda) & D(\lambda)
\end{array} \right)
=\left(
\begin{array}{cc}
{\cal A}(\lambda) & {\cal B}(\lambda)  \\
{\cal C}(\lambda) & {\cal D}(\lambda)
\end{array} \right) \label{monodromyo1} \eeq 
where the matrices
$\underline{\alpha}$, $\underline{\beta}$, $\underline{\gamma}$
and $\underline{\delta}$ are as in (\ref{K2}).

Define state $|\omega_{+}\rangle$ to be that 
with all spins up (the ferromagnetic vacuum vector): 
\be
|\omega_{+} \rangle = \underbrace{{1 \choose 0} \otimes \cdots
\otimes {1 \choose 0}}_{n+1} \,. \label{ref} \eeq  
Note that 
\be
C, C'\underbrace{{1 \choose 0} \otimes \cdots \otimes {1 \choose
0}}_{n}=0,~~\underline{\gamma}(\lambda){1 \choose 0}=0,
\label{annih} \eeq 
therefore  $|\omega_{+}\rangle$ is annihilated by ${\cal C}(\lambda)$. 
The operators ${\cal B}(\lambda)$ 
obey
\be \left[ {\cal B}(\lambda) \,, {\cal
B}(\lambda') \right] = 0 \,, \eeq 
and act as creation operators. The Bethe state 
\be |\psi \rangle ={\cal B}(\lambda_{1}) \cdots {\cal
B}(\lambda_{M})\ |\omega_{+}\rangle \label{vec2} \eeq 
is an eigenstate of the transfer matrix $t(\lambda)$, i.e. 
\be
t(\lambda)|\psi \rangle=({\cal A} +{\cal D})|\psi \rangle =
\Lambda(\lambda)|\psi \rangle\,. \label{eigenvalue} \eeq  
It is easy to determine the action of ${\cal A}$ and ${\cal D}$ on the
pseudo--vacuum (see below). 
The action of the transfer matrix on
the pseudo--vacuum, cf. (\ref{annih}), is given by
\be t(\lambda)|\omega_{+} \rangle &=& \tr_{0} \left(
\begin{array}{cc}
 A(\lambda) & B'(\lambda)  \\
 C(\lambda) & D(\lambda)
\end{array} \right)
 \left( \begin{array}{cc}
\underline{\alpha}(\lambda) & \underline{\beta}(\lambda)  \\
\underline{\gamma}(\lambda) & \underline{\delta}(\lambda)
\end{array} \right)
 \left( \begin{array}{cc}
A(\lambda) & B(\lambda)  \\
0           & D(\lambda)
\end{array} \right)|\omega_{+} \rangle \non\\&=& \Big (
\underline{\alpha}(\lambda) A^{2}
+\underline{\alpha}(\lambda)CB+\underline{\delta}(\lambda)D^{2}\Big)|\omega_{+}
\rangle \,. \label{transfer02} \eeq 
But (\ref{K2}) gives 
\be
\underline{\alpha}(\lambda){1 \choose 0}= x(\lambda;2) {1 \choose
0},~~\underline{\delta}(\lambda){1 \choose 0}=w^{+}(\lambda){1
\choose 0}, \label{vectors} \eeq 
where $x(\lambda;2)$, $w^{\pm}(\lambda)$ are given by 
(\ref{ansatz2}), (\ref{ab}). 
We have 
\be {\cal A}|\omega_{+}
\rangle=\underline{\alpha}(\lambda) A^{2}|\omega_{+}
\rangle,~~{\cal D}|\omega_{+} \rangle= \Big (
\underline{\alpha}(\lambda)CB
+\underline{\delta}(\lambda)D^{2}\Big)|\omega_{+} \rangle \eeq 
and finally,
 \be {\cal
A}|\omega_{+} \rangle = x(\lambda;2)a^{2n}(\lambda)|\omega_{+}
\rangle, ~~{\cal D}|\omega_{+} \rangle = \Big (w^{+}(\lambda)
b^{2n}(\lambda) - x(\lambda;2){a^{2n}(\lambda) -b^{2n}(\lambda)
\over a^{2}(\lambda) -b^{2}(\lambda)} \Big )|\omega_{+} \rangle.
\label{action} \eeq

 Having determined the action of the transfer matrix on the 
pseudo--vacuum, it is easy to
see via (\ref{vec2}), (\ref{eigenvalue}), (\ref{action}) that knowledge of the
commutation relations between ${\cal A}$, ${\cal B}$ and ${\cal
D}$, ${\cal B}$ is enough for the derivation of any eigenvalue. 
It is convenient \cite{Sklyanin88} to consider instead of ${\cal
D}$ the following operator 
\be \bar {\cal D} = \sinh(2\mu \lambda)
{\cal D} - \sinh(i\mu){\cal A} . \label{D} \eeq 
Then from the fundamental relation for ${\cal T}$ (\ref{bound}) it
follows that 
\be
{\cal A}(\lambda){\cal B}(\lambda_{i})=
X(\lambda,\lambda_{i}){\cal B}(\lambda_{i}){\cal
A}(\lambda)+f(\lambda, \lambda_{i}){\cal B}(\lambda){\cal
A}(\lambda_{i})+g(\lambda, \lambda_{i}){\cal B}(\lambda)\bar {\cal
D }(\lambda_{i}) \non\\  \bar {\cal D}(\lambda){\cal
B}(\lambda_{i})= Y(\lambda, \lambda_{i}){\cal B}(\lambda_{i}) \bar
{\cal D}(\lambda)+f'(\lambda, \lambda_{i}){\cal B}(\lambda){\cal
A}(\lambda_{i})+g'(\lambda, \lambda_{i}){\cal B}(\lambda)\bar
{\cal D }(\lambda_{i}), \label{cr} \eeq where \be X(\lambda,
\lambda_{i}) &=& {\sinh\mu(\lambda -\lambda_{i} -i) \over
\sinh\mu(\lambda -\lambda_{i})}{\sinh\mu(\lambda +\lambda_{i} -i)
\over \sinh\mu(\lambda +\lambda_{i})}, \non\\ Y(\lambda,
\lambda_{i}) &=& {\sinh\mu(\lambda -\lambda_{i} +i) \over
\sinh\mu(\lambda -\lambda_{i})}{\sinh\mu(\lambda +\lambda_{i}+i)
\over \sinh\mu(\lambda +\lambda_{i})} . \eeq 
The other functions
($f$, $g$, $f'$, $g'$) are not important for our purposes since
they contribute to unwanted terms, and will vanish in the
final eigenvalue expression.

We can now find the eigenvalues using the above commutation relations (\ref{cr}),
also having in mind (\ref{D}) and the action of ${\cal A}$ and  ${\cal D}$
 on the pseudo--vacuum (\ref{action}).
The eigenvalue of any Bethe ansatz state is given by
 \be \Lambda(\lambda) &=&
{\sinh \mu ( \lambda + i +i\zeta) \over \sinh (\mu i)} {\sinh \mu (
\lambda + i -i\zeta) \over \sinh (\mu i)}  \left ({\sinh \mu ( \lambda
+ i ) \over \sinh (\mu i)} \right)^{2n}{\sinh \mu (\lambda +i)
\over \sinh \mu (\lambda +{i\over 2} )} \non\\ & &
\prod_{\alpha=1}^{M} {\sinh \mu \left( \lambda - \lambda_{\alpha}
- {i\over 2} \right) \over
 \sinh  \mu \left( \lambda - \lambda_{\alpha}+{i\over 2} \right) } {\sinh \mu
 \left( \lambda +\lambda_{\alpha}
- {i\over 2} \right) \over
 \sinh  \mu \left( \lambda + \lambda_{\alpha}+{i\over 2} \right) } \non\\
& & +
 {\sinh \mu ( \lambda +i\zeta) \over \sinh (\mu i)}
{\sinh \mu ( \lambda -i\zeta) \over \sinh (\mu i)} \left( {\sinh (\mu
\lambda)  \over \sinh (\mu i)} \right)^{2n}{\sinh (\mu \lambda)
\over \sinh \mu ( \lambda + {i\over 2})} \non\\ & &
\prod_{\alpha=1}^{M} {\sinh  \mu \left( \lambda - \lambda_{\alpha}
+{3i\over 2} \right) \over
 \sinh  \mu \left( \lambda - \lambda_{\alpha} +{i\over 2}\right) }{\sinh  \mu \left( \lambda
 +\lambda_{\alpha} + {3i\over 2} \right) \over
 \sinh  \mu \left( \lambda + \lambda_{\alpha} +{i\over 2}\right) }
\,, \eeq 
provided that $\{ \lambda_{1} \,, \ldots \,, \lambda_{M}
\}$ are distinct and obey the Bethe Ansatz equations
 \be
 & &
{\sinh \mu( \lambda_{\alpha} +i\zeta +{i\over 2}) \over \sinh \mu(
\lambda_{\alpha} +i\zeta - {i\over2}) }  {\sinh \mu(\lambda_{\alpha}
-i\zeta +{i\over 2}) \over \sinh \mu ( \lambda_{\alpha} -i\zeta -
{i\over2}) }  \left( {\sinh \mu ( \lambda_{\alpha} + {i\over 2} )
\over \sinh \mu ( \lambda_{\alpha} - {i\over2}) } \right)^{2n}  =
\non\\ & & \prod_{\scriptstyle{\beta=1}\atop \scriptstyle{\beta
\ne \alpha}}^M {\sinh  \mu \left( \lambda_{\alpha} -
\lambda_{\beta} + i \right) \over
 \sinh  \mu \left( \lambda_{\alpha} - \lambda_{\beta} - i \right)}
 {\sinh  \mu \left( \lambda_{\alpha} +\lambda_{\beta} +
i \right) \over
 \sinh  \mu \left( \lambda_{\alpha} + \lambda_{\beta} - i \right) }
\,, \qquad \alpha = 1 \,, \cdots \,, M \,.
 \label{BAE1}\eeq

\section{Cabling representation}\label{cablingX}

Now consider the cabling--like representation
($\Theta: b_{n}
               \to \End(V^{2n})$) 
from \cite{MartinWoodcock01pre} discussed at the end of \S\ref{->braidgp}.  
There the elements of the blob algebra are represented as follows. 
For $U_{n\pm l} \in T_{2n}(r)$ 
let $U_{n\pm l}(r) = \calR_r(U_{n\pm l}) \in \End(V^{2n})$, 
the usual XXZ representation (\ref{u}). 
Then 
\be 
\Theta(U_l) = U_{\tilde l}(q) = U_{n-l}(r) U_{n+l}(s),
\qquad \Theta(e) = U_{\tilde 0}(Q) = {1\over i \sinh (i \mu)}U_{n}(Q)
\label{cabl}\eeq 
satisfy the relations of the blob algebra $b_n(q,\yy)$ with 
\be 
r=i\sqrt{iq}, 
~~s=\sqrt{iq},
~~Q = ie^{i \yy \mu}  \eeq 
(NB, $rs=-q$). 

Note from (\ref{cabl}) that the single index on a blob generator is
associated to a mirror image pair in the underlying $V^{2n}$. 
The $\check R$--matrix is
given by (\ref{R}), with ${\cal R}( U_{l} )=U_{\tilde l}$ 
as defined by (\ref{cabl}) 
and 
\be R_{\tilde k  \tilde l}(\lambda)= \sinh \mu(\lambda+i)
{\cal P}_{k l}{\cal P}_{k' l'}+ \sinh\mu \lambda \check U_{k l}(r)
\check U_{k' l'}(s) \label{R2} \eeq 
where we have introduced the space/mirror--space notations 
$\tilde l = (l, l')$, 
$\check U_{k l}(r)= {\cal P}_{kl}U_{k l}(r)$,
$\check R_{\tilde k \tilde l}(\lambda)={\cal P}_{\tilde k \tilde
l}R_{\tilde k \tilde l}(\lambda)$, ${\cal P}_{\tilde k \tilde
l}={\cal P}_{kl}{\cal P}_{k' l'}$. 
In the $R$--index form here, any operator
$O_{\tilde l}=O_{l l'}$ acts on $V_{l} \otimes V_{l'}$, where the
$V_{l'}$ space can be considered as the `mirror' space of
$V_{l}$ in the sense of figure~\ref{mirrorbraid1}.

This $R$--matrix satisfies the unitarity and crossing
properties 
\be R_{\tilde k \tilde l}(\lambda)R_{\tilde l \tilde
k}(-\lambda) \propto 1, 
~~R_{\tilde k \tilde l}(\lambda)=V_{\tilde
k}R_{\tilde k \tilde l}^{t_{\tilde l}}(-\lambda-i)V_{\tilde k}
\eeq 
where 
\be V_{\tilde k}=V_{k k'}= V_{k}(r)V_{k'}(s), \eeq 
and e.g.
 \be V_{k}(r)=1\otimes ... \otimes \left(
\begin{array}{cc}
                               0                  &  -ir^{{1\over 2}}  \\
                              ir^{-{1\over 2}}  & 0   \\

\end{array} \right) \otimes ... \otimes 1 \,. \eeq
This $R$--matrix is a $16 \times 16 $ matrix, 
\be R(\lambda) =
\left(
\begin{array}{cccc}
A(\lambda)     &B_{1}(\lambda) &B_{2}(\lambda) &B(\lambda)\\
C_{1}(\lambda) &A_{1}(\lambda) &B_{5}(\lambda) &B_{3}(\lambda)\\
C_{2}(\lambda) &C_{5}(\lambda) &A_{2}(\lambda) &B_{4}(\lambda)\\
C(\lambda)     &C_{3}(\lambda) &C_{4}(\lambda) &D(\lambda)\\
\end{array} \right)
\,, \label{monodromyo2} \eeq  
where the entries shown are $4 \times 4$
matrices acting on $V \otimes V$ 
(see the Appendix for the explicit form of the $R$--matrix).

The corresponding $K$--matrix (\ref{ansatz1}), (\ref{ansatz2}) is
given in matrix form by the following expression (recall that
$U_{\tilde 0}$ is given by (\ref{cabl})) 
 \be K(\lambda)=\left(
\begin{array}{cccc}
    x(\lambda;\yy)        \\
    &         w'^{-}(\lambda)  & z(\lambda) \\
    &         z(\lambda) & w'^{+}(\lambda)  \\
    &           &           & x(\lambda;\yy)

\end{array} \right) \,, \label{K3}
\eeq 
where $x(\lambda;\yy)$, $z(\lambda)$ are given by (\ref{ansatz2})
and 
\be w'^{\pm} (\lambda) &=& x(\lambda;\yy) - {1\over 2}e^{\mp
i\yy \mu}\sinh (2\mu \lambda).\eeq

The monodromy matrix has the following structure. 
\be
T_{\tilde 0}(\lambda) = \left(
\begin{array}{cccc}
{\cal A}(\lambda)     &{\cal B}_{1}(\lambda) &{\cal B}_{2}(\lambda) &{\cal B}(\lambda)\\
{\cal C}_{1}(\lambda) &{\cal A}_{1}(\lambda) &{\cal B}_{5}(\lambda) &{\cal B}_{3}(\lambda)\\
{\cal C}_{2}(\lambda) &{\cal C}_{5}(\lambda) &{\cal A}_{2}(\lambda) &{\cal B}_{4}(\lambda)\\
{\cal C}(\lambda)     &{\cal C}_{3}(\lambda) &{\cal C}_{4}(\lambda) &{\cal D}(\lambda)\\
\end{array} \right)
\,. \label{monodromyo21} \eeq 
We define a reference state 
\be
|\omega_{+}\rangle =  \underbrace{{1 \choose 0} \otimes \cdots
\otimes {1 \choose 0}}_{2n}=\otimes_{i=1}^{n}|+\rangle_{i} \,,
\label{ref1} \eeq \be |+\rangle ={1 \choose 0} \otimes  {1 \choose
0} . \eeq 
Then $C_{i}, B_{5} |+\rangle =0$, i.e.
$|\omega_{+}\rangle$ is annihilated by the operators ${\cal
C}_{i}$, ${\cal B}_{5}$. Therefore, the action of the monodromy
matrix on the reference state produces an upper triangular matrix,
\be T_{\tilde 0}(\lambda) |\omega_{+}\rangle =
\left(
\begin{array}{cccc}
{\cal A}(\lambda) &{\cal B}_{1}(\lambda) &{\cal B}_{2}(\lambda) &{\cal B}(\lambda)\\
0                 &{\cal A}_{1}(\lambda) &0                     &{\cal B}_{3}(\lambda)\\
0                 &0                     &{\cal A}_{2}(\lambda) &{\cal B}_{4}(\lambda)\\
0                 &0                     &0                     &{\cal D}(\lambda)\\
\end{array} \right) |\omega_{+}\rangle
\,. \label{monodromyo3} \eeq 
Thus for the bulk
case (\ref{monodromyo21}), (\ref{transfer}) 
the pseudo-vacuum eigenvalue is given by
\be t(\lambda)|\omega_{+}\rangle = \Big ({\cal A} + {\cal A}_{1}
+{\cal A}_{2} +{\cal D}\Big )|\omega_{+}\rangle = \Big
(a^{n}(\lambda) + b^{n}(\lambda) \Big )|\omega_{+}\rangle
\label{bulk} \eeq 
where 
\be 
{\cal A}(\lambda)=\prod_{l= 1}^{n}A^{\tilde l}, 
~~{\cal A}_{1}(\lambda)=\prod_{l= 1}^{n}A_{1}^{\tilde l}, 
~~{\cal A}_{2}(\lambda)=\prod_{l=1}^{n}A_{2}^{\tilde l}, 
~~{\cal D}(\lambda)=\prod_{ l= 1}^{n}D^{\tilde l} 
\label{def} 
\eeq 
(see also Appendix).

Now consider the open transfer matrix (\ref{transfer10}),
 \be 
t(\lambda)= tr_{\tilde 0} M_{\tilde 0} T_{\tilde 0}(\lambda) K_{\tilde 0}(\lambda)
 T_{\tilde 0}^{-1}(-\lambda), \eeq
where $K_{\tilde 0} =K_{00'}$ given by (\ref{K3}) and \be
M_{\tilde 0}= V_{\tilde 0} V_{\tilde 0}^{t} . \eeq 
Then the pseudo--vacuum eigenvalue will be 
\be
\Lambda^{0}(\lambda) &=& \langle \omega_{+}|\Big (q x(\lambda;\yy)
{\cal A}^{2}+ q^{-1} x(\lambda;\yy) {\cal D}^{2}+ q^{-1}
x(\lambda;\yy){\cal C}{\cal B} + i x(\lambda;\yy){\cal C}_{1}{\cal
B}_{1} \non\\ & &- i x(\lambda;\yy){\cal C}_{2}{\cal B}_{2} + q^{-1}
w'^{-}(\lambda){\cal C}_{3}{\cal B}_{3}+q^{-1}
w'^{+}(\lambda){\cal C}_{4}{\cal B}_{4}\Big )|\omega_{+}\rangle.
\label{a} \eeq 
where ${\cal A}$, ${\cal D}$ are given by (\ref{def}) and  
\be {\cal C}_{1,2}(\lambda)&=&\prod_{l=
1}^{n-1}A^{\tilde l}C_{1,2}^{\tilde n}, ~~{\cal
B}_{1,2}(\lambda)=\prod_{l= 1}^{n-1}A^{\tilde l}B_{1,2}^{\tilde n}
\non\\{\cal C}_{3,4}(\lambda)&=&\prod_{l=2}^{n}D^{\tilde
l}C_{3,4}^{\tilde 1}, ~~{\cal
B}_{3,4}(\lambda)=\prod_{l=2}^{n}D^{\tilde l}B_{3,4}^{\tilde 1}
\eeq 
and 
\be {\cal C}(\lambda)&=& \sum_{l= 1}^{n}D^{\tilde n}
\ldots D^{\widetilde{l+1}}C^{\tilde l}A^{\widetilde{l-1}} \ldots
A^{\tilde 1} + \sum_{l= 1}^{n-1}D^{\tilde n} \ldots
D^{\widetilde{l+2}}C_{4}^{\widetilde{l+1}}C_{2}^{\tilde
l}A^{\widetilde{l-1}} \ldots A^{\tilde 1} \non\\ &+& \sum_{l=
1}^{n-1}D^{\tilde n} \ldots
D^{\widetilde{l+2}}C_{3}^{\widetilde{l+1}}C_{1}^{\tilde
l}A^{\widetilde{l-1}} \ldots A^{\tilde 1} \eeq
 \be 
{\cal B}(\lambda)&=&
\sum_{l= 1}^{n}D^{\tilde n} \ldots D^{\widetilde{l+1}}B^{\tilde
l}A^{\widetilde{l-1}} \ldots A^{\tilde 1} + \sum_{l=
1}^{n-1}D^{\tilde n} \ldots
D^{\widetilde{l+2}}B_{4}^{\widetilde{l+1}}B_{2}^{\tilde
l}A^{\widetilde{l-1}} \ldots A^{\tilde 1} \non\\ &+& \sum_{l=
1}^{n-1}D^{\tilde n} \ldots
D^{\widetilde{l+2}}B_{3}^{\widetilde{l+1}}B_{1}^{\tilde
l}A^{\widetilde{l-1}} \ldots A^{\tilde 1}\eeq
 It is also useful to derive the action of the following operators
on the $|+ \rangle$ state:
\be 
A^{2}|+\rangle &=& a^{2}(\lambda)|+\rangle, ~~B^{2}|+\rangle =b^{2}(\lambda)|+\rangle, \non\\
C_{1}B_{1}|+\rangle &=&a^{2}(\lambda)|+\rangle, ~~C_{2}B_{2}|+\rangle=a^{2}(\lambda)|+\rangle, \non \\
CB|+\rangle&=&  \Big (a(\lambda)-q b(\lambda)\Big ) \Big (a(\lambda)-q^{-1}b(\lambda)\Big )|+\rangle, \non\\
C_{3}B_{3}|+\rangle&=& b^{2}(\lambda)|+\rangle,
~~C_{4}B_{4}|+\rangle=b^{2} (\lambda)|+\rangle. \label{b} \eeq
 Taking into account equations (\ref{a})--(\ref{b}) we conclude that the
 pseudo--vacuum eigenvalue has the form
\be
 \Lambda^{0}(\lambda) = f_{1}(\lambda)a(\lambda)^{2n}
+ f_{2}(\lambda) b(\lambda)^{2n}
 \eeq
where the functions $f_{1}(\lambda)$, $f_{2}(\lambda)$ are due to
the boundary, and are determined explicitly by
(\ref{a})--(\ref{b}) 
\footnote{ $f_{1}(\lambda)= {x(\lambda) \over
a^{2}(\lambda) -b^{2}(\lambda)} \Big \{ q  \Big (a^{2}(\lambda)
-b^{2}(\lambda) \Big)+ q^{-1} \Big (a^{2}(\lambda) +
3b^{2}(\lambda) -(q+q^{-1})a(\lambda)b(\lambda)\Big )\Big \}$

$f_{2}(\lambda) ={q^{-1} \over a^{2}(\lambda) -b^{2}(\lambda)}
\Big \{ \Big (x(\lambda) +w'^{+}(\lambda)+w'^{-}(\lambda) \Big )
\Big (a^{2}(\lambda) -b^{2}(\lambda) \Big )- x(\lambda) \Big
(3a^{2}(\lambda) +b^{2}(\lambda) -(q+q^{-1})a(\lambda)b(\lambda)
\Big ) \Big \}$}. 
The important observation here is that we are
able to derive the pseudo--vacuum eigenvalue explicitly.
Furthermore, we note that it has the expected form, compared to
the corresponding bulk eigenvalue (\ref{bulk}), in as much as the powers of
$a$ and $b$ are doubled in the open chain, and the functions
$f_{1}$ and $f_{2}$ appear as a result of the presence of the
boundary. 
The next step is the derivation of the 
general Bethe ansatz state and the corresponding eigenvalue. Here,
we do not give the details of this derivation. However  we
conjecture that the general eigenvalue will have the following
form 
\newcommand{\Agoth}{{\mathfrak A}}
\be
 \Lambda(\lambda) = f_{1}(\lambda)a(\lambda)^{2n}\Agoth_{1}(\lambda)
+ f_{2}(\lambda) b(\lambda)^{2n}\Agoth_{2}(\lambda),
 \eeq
 where $\Agoth_{1}(\lambda)$, $\Agoth_{2}(\lambda)$ can be determined
 explicitly via the algebraic or the analytical Bethe ansatz method. 
We will report on the detailed analysis of the Bethe ansatz eigenstates and
 eigenvalues, which is a separate interesting problem, in a future work.


We have arrived at this solution from abstract considerations, 
however, it clearly describes a spin chain model and it does not 
coincide with any known solution. Furthermore we have retained 
complete freedom of choice of the boundary parameter $\yy$. 
This model also has interesting symmetry properties which appear to 
significantly generalize the role of $\Uqsl2$ for ordinary XXZ. 
This makes the model a very interesting candidate for study, 
and a full spectrum analysis. 
From the representation theory of $b_n$ \cite{MartinWoodcock2000} we
know that $T_n(q)$ appears in $b_n$ in two different ways --- as a
subalgebra on dropping the boundary generator $e$, and as a quotient
for the special boundary parameter choice $\yy=1$. We also know that the
structure of $b_n$ depends profoundly on the boundary parameter
$\yy$. It will be interesting to see how the spectrum of $t(\lambda)$
depends on $\yy$, and also how the connections with $T_n(q)$ relate the
spectrum of $t(\lambda)$ here to that in the ordinary XXZ case. 
Indeed it is an interesting 
(and hopefully simpler)
preliminary question to ask what is the
spectrum 
of $t(\lambda)$  in
this `representation' {\em without} the boundary term. 
(For example, does this spectrum still depend on $r$ and $s$ separately?) 
This
should give an insight into the spectrum {\em with} boundary. 

\section{The Hamiltonian}
Here we derive the Hamiltonians of the
auxiliary string and cabling realizations.
\subsection{The auxiliary string realization}

The open spin chain Hamiltonian ${\cal H}$ is 
related to the derivative of the transfer matrix at $\lambda = 0$: 
\be {\cal H}
&=& \sum_{m=1}^{n-1} {\cal H}_{m m+1} + {1\over 4 \mu x(\lambda ;
2)}  {d\over d \lambda}
K_{1}(\lambda) \Big\vert_{\lambda=0} \non  \\
&\quad & + {\tr_{0} M_{0} {\cal H}_{n 0}\over \mu \tr M } \,,
\label{hamil01} \eeq 
where  $x(\lambda ; \yy)$ is given by (\ref{ansatz2}),
and the two--site Hamiltonian ${\cal H}_{j k}$ is given by 
\be {\cal H}_{j k} = {1 \over 2\mu} {\cal P}_{j k}
{d\over d \lambda} R_{j k}(\lambda) \Big\vert_{\lambda=0} -{1
\over 4} \cosh(i \mu) \,, \label{hamil}
 \eeq 
where the $R$--matrix is given by (\ref{R}).
This Hamiltonian is Hermitian.

Consider the model defined
by the Hamiltonian in (\ref{hamil}), (\ref{hamil01}): 
\be
{\cal H} &=& {1\over 4} \sum_{i=1}^{n-1}\Big ( \sigma_{i}^{x}
\sigma_{i+1}^{x} +\sigma_{i}^{y} \sigma_{i+1}^{y} +\cosh(i\mu)
\sigma_{i}^{z} \sigma_{i+1}^{z}\Big )+{\sinh (i \mu) \over 4}
\Big ( \sigma_{n}^{z} -\sigma_{1}^{z }  \Big ) \non\\
&+&{\sinh (i \mu) \over 4 x(0 ;2)} \Big ( \sigma_{e}^{x}
\sigma_{1}^{x} + \sigma_{e}^{y} \sigma_{1}^{y} +\cosh (i \mu)
\sigma_{e}^{z} \sigma_{1}^{z}\Big ) +{\sinh^{2} (i \mu) \over 4
x(0 ; 2)} \Big ( \sigma_{e}^{z} -\sigma_{1}^{z }  \Big )
\label{86} 
\eeq 
where $\sigma_{e}^{i}$ act on the extra space of the chain.
The bulk part is the usual XXZ
bulk spin chain with first neighbour interaction. 
The last two terms describe the boundary
interaction and come from the derivative of the 
$K$--matrix.
This Hamiltonian describes a
model which is coupled to a quantum mechanical (spin) system at
the boundaries. 
Note that it is nothing more than an $n+1$--site Hamiltonian with an
inhomogeneity at the end. 


Consider the 
Hamiltonian ${\cal H}_{f}$
obtained when we
take boundaries of the form (\ref{kd}), where $K$ is the
diagonal matrix \cite{\DVGRi,deVegaGonzalezruiz94a,deVegaGonzalezruiz93} 
\be K(\lambda) =
diag \Big (\sinh \mu (-\lambda +i \xi) e^{\mu \lambda},\ \sinh \mu
(\lambda +i \xi) e^{-\mu \lambda} \Big ) . \eeq 
By direct computation we find here 
\be 
{\cal H}_{f} & =& {\cal H} + {\coth (i\mu \xi) -1 \over 4x(0 ;
2)}
\Big ( \sinh^{2}(i \mu \zeta)\sigma_{e}^{z}  
      -\sinh^{2}(i \mu)\sigma_{1}^{z} \Big )  \non\\
& +& (\coth (i\mu \xi) -1){\sinh (i\mu) \sinh (i\mu \zeta)\over
2x(0; 2)} F_{e}(\zeta) G_{1}(-\zeta) \label{hd} 
\eeq
where ${\cal H}$ is from (\ref{86}) and 
\be
F(\zeta)=\left(
\begin{array}{cc}
                               0           & e^{i\mu {\zeta \over 2}}  \\
                               e^{-i\mu {\zeta \over 2}}   & 0   \\

\end{array} \right)\,~~ G(\zeta)=\left(
\begin{array}{cc}
                               0           & e^{i\mu {\zeta \over 2}}  \\
                               -e^{-i\mu {\zeta \over 2}}   & 0   \\

\end{array} \right)\,. \eeq
For $\zeta=0$ the above matrices become
proportional to $\sigma^{x}$ and $\sigma^{y}$ respectively.
For $i\xi\rightarrow\infty$ we see that ${\cal H}_{f}$ coincides with ${\cal H}$. 
Interestingly, the Hamiltonian ${\cal H}_{f}$ does not appear to 
have been written down explicitly before. 

\subsection{The cabling representation}

Note from (\ref{R2}) that for $\lambda =0$ the $R$--matrix reduces 
to a product of two permutation operators. 
Therefore the corresponding local Hamiltonian is defined: 
 \be 
{\cal H}_{open} &=& 
\sum_{ l= 1}^{n-1} {\cal H}_{\tilde l
                                 \mbox{\scriptsize $\widetilde{l\!\!+\!\!1}$}} 
 + {1\over 4 \mu  x(\lambda ; \yy)} {d\over d \lambda}
K_{\tilde 1}(\lambda
                     ) \Big\vert_{\lambda=0} \non\\
 &\quad &+ {\tr_{\tilde 0} M_{\tilde 0}  {\cal H}_{\tilde n \tilde 0}\over \mu \tr M} \,,
\label{hamil0} \eeq
 where the two--site Hamiltonian ${\cal H}_{\tilde k \tilde l}$
is given by 
\be {\cal H}_{\tilde k \tilde l} = {1 \over 2\mu}
{\cal P}_{\tilde k \tilde l} {d\over d \lambda} R_{\tilde k \tilde
l}(\lambda) \Big\vert_{\lambda=0} -{1 \over 4} \cosh(i\mu) \,,
\label{hamil1}
 \eeq
and the $R$--matrix is given by (\ref{R2}). 
Unlike (\ref{86}) 
this is completely new. 
It will be studied in detail elsewhere. 

\[ \]
{\bf Acknowledgements.} PPM would like to thank EPSRC for partial
financial support under GRM22536. AD would like to thank EPSRC for 
a research fellowship, and P Baseilhac, 
G Delius and V Kuznetsov for useful conversations. 
\appendix
\section*{Appendix}
In this section we write explicitly the $16 \times 16$ $R_{\tilde
k \tilde l}$ matrix. In particular we write down the $4\times 4$
entries of the matrix, \be  A(\lambda) = \left(
\begin{array}{cccc}
 a(\lambda)    &0 &0 &0\\
0 &0 &0 &0\\
 0          &0 &0 &0\\
 0          &0 &0 &b(\lambda)\\
\end{array} \right)
\,,~~ D(\lambda) = \left(
\begin{array}{cccc}
 b(\lambda)   &0 &0 &0\\
 0  &0 &0 &0\\
 0          &0 &0 &0\\
 0          &0 &0 &a(\lambda)\\
\end{array} \right)
\,, \eeq \be  A_{1}(\lambda) = \left(
\begin{array}{cccc}
 0    &0 &0 &0\\
0 &a(\lambda) &0 &0\\
 0          &0 &b(\lambda) &0\\
 0          &0 &0 &0\\
\end{array} \right)
\,,~~ A_{2}(\lambda) = \left(
\begin{array}{cccc}
 0   &0 &0 &0\\
 0  &b(\lambda) &0 &0\\
 0          &0 &a(\lambda) &0\\
 0          &0 &0 &0\\
\end{array} \right)
\,, \eeq \be  B_{1}(\lambda) = \left(
\begin{array}{cccc}
 0    &0 &0 &0\\
a(\lambda) &0 &0 &0\\
 0          &0 &0 &0\\
 0          &0 &-s b(\lambda) &0\\
\end{array} \right)
\,,~~ B_{2}(\lambda) = \left(
\begin{array}{cccc}
 0    &0 &0 &0\\
 0  &0 &0 &0\\
 a(\lambda)          &0 &0 &0\\
 0          &-r b(\lambda) &0 &0\\
\end{array} \right)
\,, \eeq \be B_{5}(\lambda) = \left(
\begin{array}{cccc}
 0    &0 &0 &0\\
 0  &0 &0 &0\\
 0          &a(\lambda)-rs^{-1} b(\lambda) &0 &0\\
 0          &0 &0 &0\\
\end{array} \right)
\,, ~~B(\lambda) = \left(
\begin{array}{cccc}
 0    &0 &0 &0\\
 0  &0 &0 &0\\
 0          &0 &0 &0\\
 a(\lambda)-q b(\lambda)          &0 &0 &0\\
\end{array} \right)
\, \eeq  $B_{3}$, $B_{4}$ have the same structure as $B_{2}$,
$B_{1}$ respectively, with the matrix elements interchanged. 
\newcommand{\pxi}{p}%
Also,
$C_{i}(\pxi) = B_{i}(\pxi^{-1})^{t}$, where $\pxi$ is in
general the anisotropy parameter, it can be $r$, $s$, $q$.




\bibliographystyle{amsplain}
\bibliography{new31,main,emma}

\providecommand{\bysame}{\leavevmode\hbox to3em{\hrulefill}\thinspace}
\providecommand{\MR}{\relax\ifhmode\unskip\space\fi MR }
\providecommand{\MRhref}[2]{%
  \href{http://www.ams.org/mathscinet-getitem?mr=#1}{#2}
}
\providecommand{\href}[2]{#2}
\begin{thebibliography}{10}

\bibitem{AffleckOshikawaSaleur98}
I~Affleck, M~Oshikawa, and H~Saleur, \emph{Boundary critical phenomena in the
  three--state {P}otts model}, cond-mat/9804117 (1998).

\bibitem{Allcock99}
D~Allcock, \emph{Braid pictures for {A}rtin groups}, math.GT/9907194 (1999).

\bibitem{BaseilhacDelius01}
P~Baseilhac and G~Delius, J Phys \textbf{A34} (2001), 8259.

\bibitem{Baxter82}
R~J Baxter, \emph{Exactly solved models in statistical mechanics}, Academic
  Press, New York, 1982.

\bibitem{BaxterKellandWu76}
R~J Baxter, S~B Kelland, and F~Y Wu, \emph{Equivalence of the {P}otts model or
  {W}hitney polynomial with an ice--type model}, J Phys A \textbf{9} (1976),
  397--406.

\bibitem{BaxterTemperleyAshley78}
R~J Baxter, H~N~V Temperley, and S~E Ashley, Proc R Soc A \textbf{358} (1978),
  535.

\bibitem{BilsteinRittenberg00}
U~Bilstein and V~Rittenberg, cond-mat/0004241 (2000).

\bibitem{Brieskorn73}
E~Brieskorn, \emph{Sur les groupes des tresses [d'apres {V.}{I.}{Arnol'd}]},
  Seminaire {B}ourbaki, Exp. No. 401, Lec {N}otes in {M}ath, no. 317, Springer,
  1973, pp.~21--44.

\bibitem{BrieskornSaito72}
E~Brieskorn and K~Saito, \emph{Artin--gruppen und {C}oxeter--gruppen},
  Inventiones math. \textbf{17} (1972), 245--271, N.B., there exists a
  translation into English by Coleman, Corran, Crisp, Easdown, Howlett, Jackson
  and Ram (Sydney 1996).

\bibitem{CastroFringGohmann02}
O~A Castro-Alvaredo, A~Fring, and F~Gohmann, hep-th/0201142 (2002).

\bibitem{Cherednik84}
I~Cherednik, Theor Math Phys \textbf{61} (1984), 977.

\bibitem{Cherednik91}
\bysame, \emph{A unification of {K}nizhnik--{Z}amolodchikov and {Dunkl}
  operators via affine {H}ecke algebras}, Invent Math \textbf{106} (1991),
  411--431.

\bibitem{deVega89}
H~J de~Vega, \emph{Yang-{B}axter algebras, integrable theories and quantum
  groups}, Int J Mod Phys A \textbf{4} (1989), 2371--2463.

\bibitem{deVegaGonzalezruiz93}
H~J de~Vega and A~Gonz\'alez-Ruiz, J Phys A \textbf{26} (1993), L519.

\bibitem{deVegaGonzalezruiz94c}
\bysame, Nucl Phys \textbf{B417} (1994), 553.

\bibitem{deVegaGonzalezruiz94a}
\bysame, \emph{Exact {B}ethe-ansatz solution for a(n-1) chains with
  non-$su_q(n)$ invariant open boundary-conditions}, Modern Physics Letters A
  \textbf{9} (1994), 2207--2216.

\bibitem{DipperJamesMathas99}
R~Dipper, G~James, and A~Mathas, \emph{The $({Q},q)$--{S}chur algebra},
  preprint (1999).

\bibitem{DoikouNepomechie98}
A~Doikou and R~I Nepomechie, Nucl Phys \textbf{B530} (1998), 641.

\bibitem{FaddeevTakhtajan79}
L~D Faddeev and L~A Takhtajan, Russ Math Surv \textbf{34} (1979), 11.

\bibitem{FaddeevTakhtajan81}
\bysame, Phys Lett \textbf{85A} (1981), 375.

\bibitem{Fendley93}
P~Fendley, Phys Rev Lett \textbf{71} (1993), 2485.

\bibitem{FrahmSlavnov99}
H~Frahm and N~A Slavnov, J Phys \textbf{A32} (1999), 1547.

\bibitem{FringKoberle94a}
A~Fring and R~Koberle, Nucl Phys \textbf{B421} (1994), 159.

\bibitem{FringKoberle94b}
\bysame, Nucl Phys \textbf{B419} (1994), 647.

\bibitem{GeGouldLinksZhou99}
X-Y Ge, M~D Gould, J~Links, and H-Q Zhou, cond-mat/9908191 (1999).

\bibitem{GhoshalZamolodchikov94a}
S~Ghoshal and A~B Zamolodchikov, Int J Mod Phys A \textbf{9} (1994), 3841.

\bibitem{Henkel99}
M~Henkel, \emph{Conformal invariance and critical phenomena}, Texts and
  monographs in {P}hysics, Springer, 1999.

\bibitem{Humphreys90}
J~E Humphreys, \emph{Reflection groups and {C}oxeter groups}, Cambridge
  University Press, 1990.

\bibitem{Jones94b}
V~F~R Jones, \emph{A quotient of the affine {H}ecke algebra in the {B}rauer
  algebra}, L'Enseignement Math\'ematique \textbf{40} (1994), 313--344.

\bibitem{Joseph95}
A~Joseph, \emph{Quantum groups and their primitive ideals}, Springer--Verlag,
  1995.

\bibitem{KashiwaraMiwaStern95}
M~Kashiwara, T~Miwa, and E~Stern, \emph{Decomposition of $q$--deformed {F}ock
  spaces}, q-alg/9508006, 1995.

\bibitem{KazhdanLusztig79}
D~Kazhdan and G~Lusztig, \emph{Representations of coxeter groups and {H}ecke
  algebras}, Inventiones Math. \textbf{53} (1979), 165--184.

\bibitem{KooSaleur93}
W~M Koo and H~Saleur, Int J Mod Phys A \textbf{8} (1993), 5165--5233.

\bibitem{KorepinIzerginBogoliubov93}
V~Korepin, G~Izergin, and N~M Bogoliubov, \emph{Quantum inverse scattering
  method, correlation functions, and algebraic {B}ethe ansatz}, Cambridge UP,
  1993.

\bibitem{KulishSklyanin91}
P~P Kulish and E~K Sklyanin, \emph{The general $u_q[sl(2)]$ invariant {XXZ}
  integrable quantum spin chain}, J Phys A \textbf{24} (1991), L435--L439.

\bibitem{Kuznetsov01}
V~Kuznetsov, private communication (2001).

\bibitem{LevyMartin94}
D~Levy and P~Martin, \emph{Hecke algebra solutions to the reflection equation},
  J Phys A \textbf{27} (1994), L521--L526.

\bibitem{Martin89b}
P~P Martin, \emph{Block spin transformations in the operator formulation of
  two--dimensional {P}otts models}, J Phys A \textbf{22} (1989), 3991--4005.

\bibitem{Martin91}
\bysame, \emph{Potts models and related problems in statistical mechanics},
  World Scientific, Singapore, 1991.

\bibitem{Martin92}
\bysame, \emph{On {S}chur-{W}eyl duality, {$A_n$} {H}ecke algebras and quantum
  {$sl(N)$}}, Int J Mod Phys A \textbf{7 suppl.1B} (1992), 645--674.

\bibitem{MartinSaleur93}
P~P Martin and H~Saleur, \emph{On an algebraic approach to higher dimensional
  statistical mechanics}, Commun. Math. Phys. (1993), no.~158, 155--190.

\bibitem{MartinSaleur94a}
\bysame, \emph{The blob algebra and the periodic {T}emperley--{L}ieb algebra},
  Lett. Math. Phys. (1994), no.~30, 189--206.

\bibitem{MartinSaleur94c}
\bysame, \emph{On algebraic diagonalisation of the {XXZ} chain}, Int J Mod Phys
  B \textbf{8} (1994), 3637--3644.

\bibitem{MartinWoodcock01pre}
P~P Martin and D~Woodcock, \emph{On the blob algebra and generalisations},
  Tech. report, City University, 1998, also available as {\em Generalized blob
  algebras and alcove geometry}, math.RT/0205263.

\bibitem{MartinWoodcock2000}
\bysame, \emph{On the structure of the blob algebra}, J Algebra \textbf{225}
  (2000), 957--988.

\bibitem{MartinWoodcockLevy00}
P~P Martin, D~Woodcock, and D~Levy, \emph{A diagrammatic approach to {H}ecke
  algebras of the reflection equation}, J Phys A \textbf{33} (2000),
  1265--1296.

\bibitem{MezincescuNepomechie91a}
L.~Mezincescu and R.~Nepomechie, J. Phys. A: Math. Gen. \textbf{24} (1991),
  L17--L23.

\bibitem{MezincescuNepomechie91c}
\bysame, Mod Phys Lett \textbf{A6} (1991), 2497.

\bibitem{MezincescuNepomechie92}
\bysame, Int. J. Mod. Phys. A \textbf{7} (1992), 5657.

\bibitem{PasquierSaleur90}
V~Pasquier and H~Saleur, Nucl Phys B \textbf{330} (1990), 523.

\bibitem{Reidemeister48}
K~Reidemeister, \emph{Knotentheorie}, Chelsea, New York, 1948.

\bibitem{ReshetikhinSemenovTianShansky90}
N~Yu Reshetikhin and M~A Semenov-Tian-Shansky, Lett Math Phys \textbf{19}
  (1990), 133.

\bibitem{RochaCaridi84}
A~Rocha-Caridi, \emph{Vacuum vector representations of the {V}irasoro algebra},
  in {V}ertex Operators in Mathematics and Physics (MSRI \#3, Springer) (1984),
  451--473.

\bibitem{Sklyanin88}
E~K Sklyanin, \emph{Boundary conditions for integrable quantum systems}, J Phys
  A \textbf{21} (1988), 2375--2389.

\bibitem{TemperleyLieb71}
H~N~V Temperley and E~H Lieb, Proceedings of the Royal Society A \textbf{322}
  (1971), 251--280.

\bibitem{TsvelickWiegmann83}
A~M Tsvelick and P~B Wiegmann, Adv in Phys \textbf{32} (1983), 453.

\end{thebibliography}
\end{document}